\documentclass{aa}  
\usepackage{natbib}
\bibpunct{(}{)}{;}{a}{}{,} 
\usepackage{graphicx}
\usepackage{txfonts}
\usepackage{subfig}
\usepackage{soul}
\usepackage[dvipsnames]{xcolor}
\usepackage{amsmath}
\usepackage{hyperref}
\hypersetup{
    colorlinks=true,
    citecolor=blue,
    linkcolor=blue,
    filecolor=magenta,      
    urlcolor=cyan}
\usepackage{caption}
\usepackage{lipsum}
\usepackage{animate}
\usepackage[normalem]{ulem}
\usepackage{multirow}   
\usepackage{makecell}
\usepackage{booktabs,tabularx}
\newcommand{\xmm}{{\it XMM-Newton}}

\usepackage{placeins}

\begin{document}

\title{Fooled by mass? Recovering the $\Gamma-\lambda_\textup{Edd}$ relation of quasars from their X-ray variability}

  \author{Matilde Signorini\inst{1,2}\thanks{\email{matilde.signorini@esa.int}} Andrea Sacchi\inst{3,4}, Bartolomeo Trefoloni\inst{2,5}, Matteo Guainazzi \inst{1}, Elisabeta Lusso\inst{2,6}, Emanuele Nardini\inst{2}, Chiara Niccolai\inst{7,8}, Federica Ricci\inst{9}, Anastasia Shlentsova\inst{6,10}, Mateusz Ra{\l}owski\inst{2,6,11}}
  
   \institute{European Space Agency (ESA), European Space Research and Technology Centre (ESTEC), Keplerlaan 1, 2201 AZ Noordwijk, the Netherlands
\and INAF - Osservatorio Astrofisico di Arcetri, Largo Enrico Fermi 5, I-50125 Firenze, Italy
\and INAF - Istituto di Astrofisica Spaziale e Fisica Cosmica di Milano, via A. Corti 12, I-20133 Milano, Italy
\and Center for Astrophysics $\vert$ Harvard \& Smithsonian, 60 Garden Street, Cambridge, MA 02138, USA
\and Scuola Normale Superiore, Piazza dei Cavalieri 7, I-56126 Pisa, Italy
\and Dipartimento di Fisica e Astronomia, Università degli Studi di Firenze, via G. Sansone 1, 50019 Sesto Fiorentino, Firenze, Italy
\and Scuola Superiore Meridionale, Via Mezzocannone 4, I-80138 Napoli, Italy
\and Istituto Nazionale di Fisica Nucleare, sez. di Napoli, Via Cinthia 9, I-80126 Napoli, Italy
\and Dipartimento di Matematica e Fisica, Università Roma Tre, via della Vasca Navale 84, I-00146, Roma, Italy
\and Instituto de Astrofísica, Facultad de Física, Pontificia Universidad Católica de Chile, Casilla 306, Santiago 22, Chile
\and Astronomical Observatory of the Jagiellonian University, Faculty of Physics, Astronomy and Applied Computer Science, ul. Orla 171, 30-244 Cracow, Poland
    }

\titlerunning{Recovering the $\Gamma-\lambda_\textup{Edd}$ relation}
\authorrunning{M. Signorini et al. 2026}

   \date{\today}

  \abstract
{We measure the relation between the X-ray photon index $\Gamma$ and the Eddington ratio $\lambda_{\rm Edd}$ in unobscured type-1 quasars, using as parent sample the cross-match of the SDSS DR16Q catalogue with the 4XMM-DR14 serendipitous catalogue. After cleaning the sample of potential observational biases, we obtain 7835 objects, one order of magnitude larger than most previous analyses in the literature. A single regression of $\Gamma$ against $\log\lambda_{\rm Edd}$, with single-epoch virial black-hole masses, returns slopes that depend on the mass indicator (H$\beta$, Mg\,{\sc ii}, C\,{\sc iv}) and on the bolometric correction, and that are generally flat. We interpret this as regression dilution: the systematic uncertainty on virial masses is much larger than the intrinsic spread of $\lambda_{\rm Edd}$ in SDSS quasars.  We therefore introduce a different approach, which removes the black-hole mass from the problem. For quasars observed by {\it XMM-Newton} at multiple epochs, the mass of each object is fixed, and changes in its X-ray luminosity trace changes in its accretion state. We fit the $\Gamma$--$\log L_{2\,\rm keV}$ relation of each object (for a total sample of N=893) and combine the per-object posteriors. This recovers the mean response of the population $\langle a_X\rangle = 0.35\pm0.03$, which steepens to $\approx0.43$ when only epochs with precisely measured photon indices are used. Translated to the Eddington ratio, this corresponds to $\langle a_\lambda\rangle\approx0.20$--0.25, depending on the choice of bolometric correction and on the $L_{\rm X}$--$L_{\rm UV}$ slope. The $\Gamma$--$\lambda_{\rm Edd}$ relation therefore exists, but it can be uncovered only by exploiting the variability of individual sources, which removes the black-hole mass from the problem. Moreover, its shallowness implies that the differences in Eddington ratio, as currently measured, cannot by themselves account for the variety of photon indices observed among unobscured quasars.
}


\maketitle

\section{Introduction}
\label{sec:intro}
The energy released by accretion onto a supermassive black hole (SMBH) in active galactic nuclei (AGN) emerges across the full electromagnetic spectrum. In unobscured type-1 sources two components carry the observational signatures of the innermost accretion flow: a thermal ``big blue bump'' from the accretion disc, peaking in the ultraviolet, and a power-law X-ray continuum produced by inverse Compton scattering of disc photons in a hot, optically thin electron plasma called X-ray corona \citep{Haardt91, Haardt93}. The existence of this compact corona is firmly established by X-ray variability and microlensing studies \citep[e.g.,][]{Chartas09, Reis13, Jiang19} as well as, more recently, by polarization detections with the Imaging X-ray Polarimetry Explorer (IXPE) \citep{Marinucci22, Gianolli23}. However, the mechanism by which gravitational energy is extracted from the disc and conveyed to the corona (e.g., magnetic reconnection \citep{Galeev79, Liu02}, vertical advection \citep{Merloni01}) remains debated, as does its size and geometry. Polarimetry measurements of Type-1 Seyferts favour a slab- or wedge-like corona aligned with the radio axis and disfavour a simple spherical lamp-post configuration \citep[e.g.,][]{Marinucci22, Tagliacozzo23, Pal23}, but the sample size of these studies is still limited.\\
A direct empirical probe of the disc--corona coupling is the X-ray photon index $\Gamma$, which encodes the temperature and optical depth of the corona \citep[e.g.,][]{Zdziarski85, Middei19} and should therefore respond to changes in the seed photon field from the disc. Standard accretion theory predicts $\Gamma$ to correlate positively with the Eddington ratio $\lambda_{\rm Edd}$\,$\equiv$\,$L_{\rm bol}/L_{\rm Edd}$: a larger disc contribution cools the corona more efficiently, steepening the X-ray spectrum \citep[e.g.,][]{Done12} and resulting in a diagnostic of the accretion state analogous to that used in X-ray binaries \citep{Done07, Sobolewska09}. This `softer-when-brighter' behaviour has indeed been detected both within individual AGN tracked across flux states \citep[e.g.,][]{SobolewskaPapadakis09} and as a population trend.
Observationally, the $\Gamma$--$\lambda_{\rm Edd}$ relation has been investigated for over two decades, but the picture is still fragmented. \cite{Shemmer06} first showed that $\lambda_{\rm Edd}$, rather than $M_{\rm BH}$ or FWHM(H$\beta$) alone, is connected to the coronal conditions in Palomar--Green quasars. Using 35 radio-quiet AGN with H$\beta$-based masses, \citet[][hereafter S08]{Shemmer08} found $\Gamma=(0.31\pm0.01)\log\lambda_{\rm Edd}+2.11$. \citet[][R09]{Risaliti09} analysed $\sim$\,350 SDSS quasars with serendipitous \xmm{} spectra and found a slope of $0.31\pm0.06$ for the whole sample, steeper for objects with H$\beta$-based masses ($0.58\pm0.11$), weaker for Mg\,{\sc ii} ($0.27\pm0.09$), and no correlation for C\,{\sc iv}. \citet[][B13]{Brightman13} obtained $0.32\pm0.05$ for 69 X-ray bright AGN in COSMOS and E-CDF-S with H$\alpha$- and Mg\,{\sc ii}-based masses. In these works $L_{\rm bol}$ was derived from the optical/UV continuum (S08, R09) or from SED fitting, with a bolometric correction to the 2--10\,keV luminosity for the part of the B13 sample without SED data. \cite{Kelly08} and \cite{Fanali13} broadly confirmed a positive trend, intertwined with a modest $\Gamma$--$M_{\rm BH}$ anti-correlation. A more complicated picture emerged with \citet[][T17]{Trakhtenbrot17}, who applied broadband ($\sim$\,0.3--150 keV) spectral modelling to 228 hard-X-ray-selected AGN from the BAT AGN Spectroscopic Survey \citep[BASS,][]{Koss17}. With $L_{\rm bol}$ derived mainly from the 2--10\,keV luminosity, they found a significant but weak correlation, $\Gamma=(0.167\pm0.04)\log\lambda_{\rm Edd}+2.00$, and no significant correlation within the subsets of AGN with direct or single-epoch masses. A further BASS study \citep{Ricci18} found the coronal high-energy cutoff $E_{\rm C}$ to be anti-correlated with $\lambda_{\rm Edd}$ (computed from the intrinsic 2--10\,keV luminosity with a constant bolometric correction), consistent with cooler, more compact coronae at high accretion rates.

This picture is further complicated in the high-$\lambda_{\rm Edd}$ regime, where the canonical (slope $\sim$0.3) relation would predict $\Gamma$\,$\gtrsim$\,2. \cite{Laurenti22} found that 14 radio-quiet quasars with $\lambda_{\rm Edd}\gtrsim1$ and $M_{\rm BH}\sim10^{8}$--$10^{8.5}\,M_\odot$ span $\Gamma$\,=\,1.3--2.5, and that $\sim$\,30\% of them are X-ray weak by a factor of 10--80 with respect to typical quasars of similar UV luminosity, apparently intrinsically rather than because of absorption. A comparably large spread in photon indices, and a high incidence of X-ray weak sources, was found in the brightest blue quasars at $z$\,$\sim$\,3 \citep{Nardini19, Trefoloni23}, for which \cite{Trefoloni23} measured a flat $\Gamma$--$\lambda_{\rm Edd}$ slope of $0.16\pm0.03$. \cite{Liu21} reported a slope of $0.27\pm0.04$ for 47 local reverberation-mapped AGN, intermediate between T17 and the canonical value. \cite{Laurenti24} added 61 highly accreting AGN ($-0.2<\log\lambda_{\rm Edd}<0.5$) from the X-HESS sample: they found a large scatter and no significant correlation at high $\lambda_{\rm Edd}$, and a slope of $0.23\pm0.04$, with a dispersion of 0.22 in $\Gamma$, only when combining their sample with that of \cite{Liu21}. This supports the view that the scatter of $\Gamma$ at fixed $\lambda_{\rm Edd}$ is intrinsic, and reflects diverse coronal properties rather than a unique disc--corona response.

\begin{figure}[!htbp]
\centering
\includegraphics[width=\linewidth,clip]{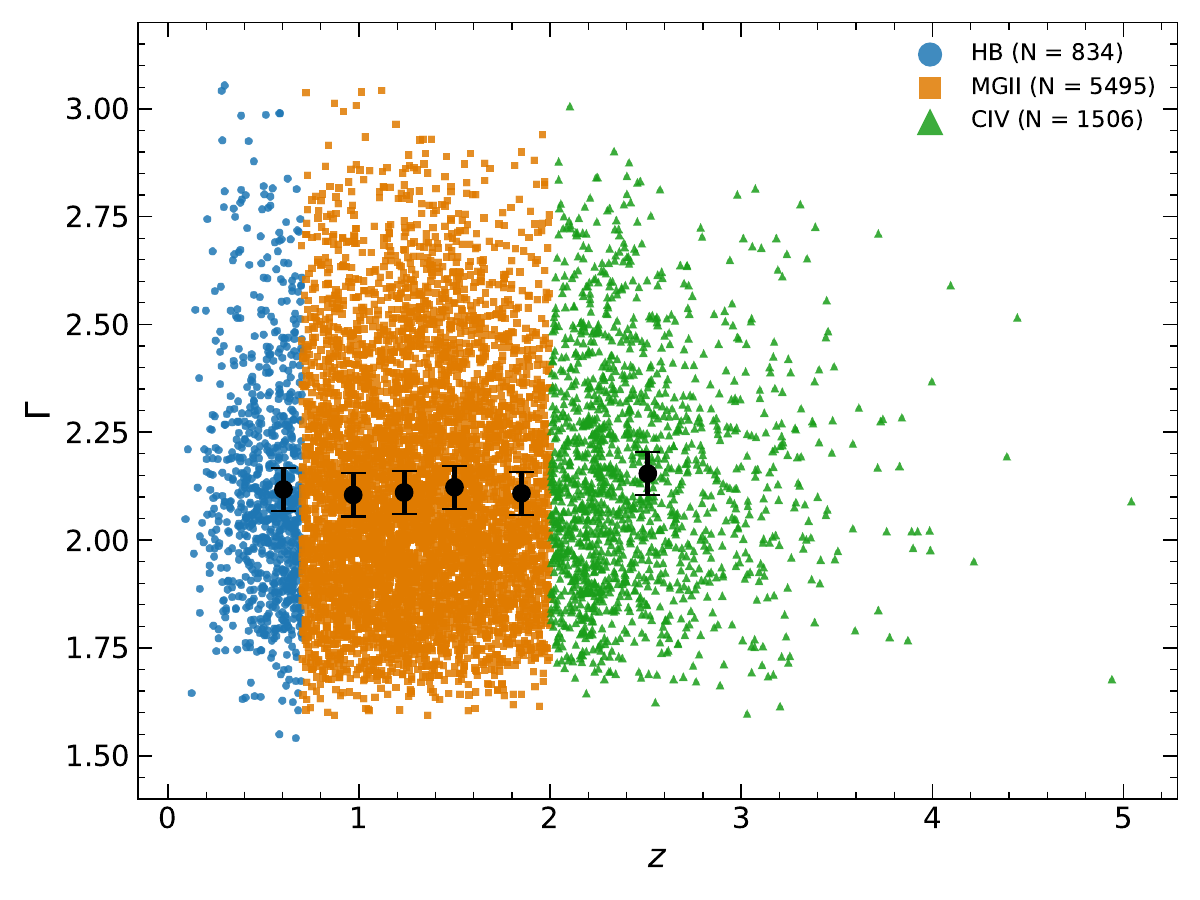}
\includegraphics[width=\linewidth,clip]{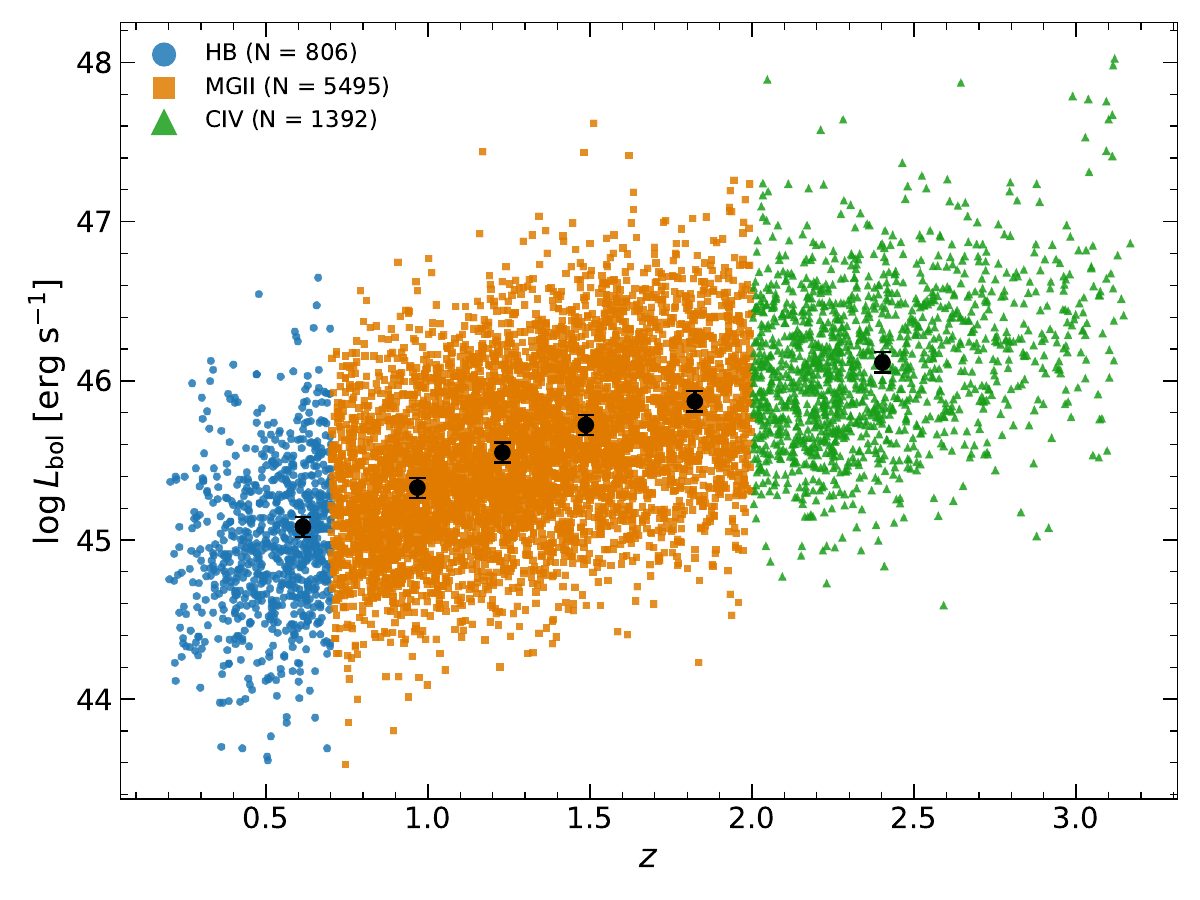}
\caption{Upper: Photon index as a function of redshift for the clean sample. Different colors and shapes indicate different broad lines used for the SE $M_{BH}$ measurements. Black dots show the mean of $\Gamma$ in redshift bins. There is no significant trend with the redshift. Lower: bolometric luminosity (K13) as a function of redshift for the clean sample.}
\label{fig:gamma_z}
\end{figure}

The uncertainties affecting this relation become critical when one aims to extend our understanding of the disc--corona connection to high redshift: the populations of high-redshift AGN recently discovered by JWST observations, including the so-called little red dots, reveal accreting black holes that appear remarkably X-ray weak relative to local AGN \citep[e.g.][]{Ananna24, Yue24, Maiolino25, Sacchi25, Brazzini26}. This is interpreted in some of these studies as a signature of high-Eddington accretion, with very steep X-ray spectra: a puzzle for which a robust local benchmark of the disc--corona connection is needed. We note that high-Eddington accretion is only one possible interpretation; the X-ray weakness could also be due to extreme obscuration and/or to a corona that is not properly developed \citep{Maiolino25, Sacchi25, Madau24}.

In this paper we investigate the $\Gamma$--$\lambda_{\rm Edd}$ relation using the largest homogeneous sample assembled for this purpose to date: the cross-match of the SDSS DR16 quasar catalogue \citep{Wu22} with the \xmm{} serendipitous source catalogue \citep{Webb20}, selected to minimise dust reddening, host contamination, and X-ray obscuration (following \citealt{Lusso20}), yielding 7835 type-1 quasars at $z$\,=\,0--7. We first show that a regression of $\Gamma$ against $\log\lambda_{\rm Edd}$ built on single-epoch virial masses cannot measure the relation reliably. We then use the objects observed at multiple epochs to measure the response of $\Gamma$ at fixed black-hole mass, and discuss what the resulting, rather shallow, relation implies for the physical role of the Eddington ratio.

The paper is structured as follows. Section~\ref{sec:sample} describes the sample selection and X-ray photometric measurements. Section~\ref{sec:eddratio} presents the bolometric corrections and black-hole masses. Section~\ref{sec:wholesample} presents the whole-sample analysis. Section~\ref{sec:multiepoch} presents the multi-epoch per-object analysis. Section~\ref{sec:discussion} discusses the implications, and Sect.~\ref{sec:conclusions} summarises our results. We adopt a flat $\Lambda$CDM cosmology with $H_0=70$ km s$^{-1}$ Mpc$^{-1}$ and $\Omega_M=0.3$; uncertainties are quoted at the 68\,\% confidence level.

\section{Sample}
\label{sec:sample}

The parent sample we start from is the Sloan Digital Sky Survey (SDSS) DR16 quasar catalogue \citep{Wu22}, which contains 750,414 spectroscopically confirmed broad line AGN. From that, we exclude broad absorption line (BAL) quasars (BI\_CIV\,$>$\,0 in the SDSS catalogue) and radio-loud sources, since both populations have UV and X-ray emission that may be contaminated by absorption (BAL) or by a jet component (RL). Following \cite{Lusso20}, we compute the radio loudness $R\equiv L_{\nu,6\,\rm cm}/L_{\nu,2500\,\rm \AA}$ for the quasars detected by FIRST and classify as radio loud those with $R>10$; we also exclude the sources flagged as radio loud in the MIXR catalogue \citep{Mingo16}. We refer to \cite{Lusso20} for details.
We cross-match the resulting catalogue with the \xmm{} serendipitous source catalogue \citep[4XMM--DR14;][]{Webb20}, which contains 692,109 unique X-ray sources, adopting a maximum separation of 3 arcsec. We performed a cleaning procedure in order to keep only blue, broad-line quasars with reliable X-ray and UV fluxes. This procedure has been described and adopted in several works \citep{Lusso16,RL19_nature,Lusso20, Risaliti26b}, so here we only summarise the main steps.
First, we applied the following quality cuts in the 4XMM--DR14 catalogue: SUM\_FLAG\,$<$\,3 (low level of spurious detections), OBS\_CLASS\,$\leq$\,3 (quality classification of the whole observation), EP\_TIME\,$>$\,0 (EPIC exposure time available) and CONFUSED\,=\,False. From the SDSS side, we keep all spectroscopically confirmed broad-line quasars with redshift 0\,$<$\,$z$\,$<$\,7 for which a single-epoch black-hole mass from either the H$\beta$, the Mg\,{\sc ii}, or the C\,{\sc iv} emission line is available \citep{shen13, Wu22}.
This sample results in 18252 objects, for a total of 28756 detections. From this, we remove objects whose UV and/or X-ray fluxes might be underestimated due to dust reddening and/or gas absorption. The number of sources left after each of the steps described below is reported in Table~\ref{tab:cutflow}.

We select quasars whose optical/UV continuum is consistent with an unobscured type-1 AGN. For every source, we construct a photometric spectral energy distribution (SED) from the SDSS, GALEX, 2MASS, and WISE bands as described in \citet[][their Section 7]{Lusso20}, and we measure the slopes of the $\log\nu$--$\log(\nu L_\nu)$ power law in the rest-frame 0.3--1\,$\mu$m ($\alpha_1$) and 1450--3000\,\AA\ ($\alpha_2$) ranges. Dust-reddened sources occupy a well-defined region of the ($\alpha_1$,$\alpha_2$) plane \citep[see also][]{Hao13}, displaced from the locus of intrinsically blue quasars \citep{Richards06, Lusso20}. We retain only quasars within a circle of radius corresponding to $E(B-V)\lesssim 0.1$ around $(\alpha_1,\alpha_2)$\,=\,(0.82,0.40), applying an SMC reddening law \citep{Prevot84}; the distribution of the parent sample in this plane and the adopted selection are shown in \citet[][their Fig.~1]{Risaliti26b}. This cut removes red quasars, sources with strong host-galaxy contamination at 1\,$\mu$m, and a small number of objects with unphysical SEDs (with likely defective photometry).

To compute the rest-frame 2-keV monochromatic luminosity and the photon index $\Gamma$ from the 4XMM--DR14 broadband measurements, we follow the procedure described by \cite{RL19_nature} and \cite{Lusso20}. In brief, for each observation we consider the tabulated 1--2~keV and 2--4.5~keV fluxes reported in the 4XMM--DR14 catalogue. We identify the pivot energies for each band, which are the energies at which the covariance between the band flux and the assumed power-law slope is zero; the monochromatic fluxes at these pivot energies are then, by construction, insensitive to the assumed $\Gamma$ used to derive band-integrated 4XMM fluxes. The photometric photon index is obtained as the slope of the power law connecting the two pivot-energy fluxes, and the rest-frame 2-keV flux is interpolated (or extrapolated) along this same power law. We use the 1--4.5 keV \xmm\ band rather than the standard soft (0.5--2 keV) and hard (4--12 keV) bands, as the former can be more strongly contaminated by the soft excess \citep[e.g.,][]{Sobolewska07} at low redshift ($z \lesssim 1$), while the latter is noise-dominated for the serendipitous sources that constitute the bulk of the sample. For more details, we refer to Section 2.2 of \cite{Risaliti26b}. We denote with $\Delta\Gamma$ the 1$\sigma$ uncertainty on the photometric $\Gamma$, propagated from the uncertainties on the two band fluxes.

Finally, we require $\Gamma - \Delta\Gamma>1.5$ and $\Gamma + \Delta\Gamma  < 3.5$, where $\Delta\Gamma$ is the 1-$\sigma$ uncertainty on the photon index $\Gamma$. The upper limit only removes a handful of objects with unphysically steep photometric slopes; we verified that lowering it to 3 does not change any of the results, as only 10 sources lie above this value. The lower limit is designed to exclude sources with significant X-ray obscuration, and deserves some discussion, as it truncates the low end of the $\Gamma$ distribution and could in principle flatten the recovered $\Gamma$--$\lambda_{\rm Edd}$ slope. On physical grounds, very flat intrinsic spectra are hard to obtain in thermal comptonisation models of radiatively efficient coronae, where Compton cooling by the disc photons keeps $\Gamma\approx2$ unless the corona is strongly photon-starved \citep{Haardt93, Haardt94, Stern95}. Observationally, large samples of unobscured type-1 AGN consistently give $\langle\Gamma\rangle\approx1.8$--2.0, both for optically selected quasars with serendipitous \xmm{} spectra \citep[$\langle\Gamma\rangle=1.91\pm0.08$ and $1.99\pm0.01$;][]{Young09, Scott11} and for local, hard-X-ray-selected AGN (median $\Gamma=1.77$; \cite{Gupta24}).
Given the photometric nature of our $\Gamma$ measurements, mild obscuration ($N_{\rm H}$\,$\sim$\,$10^{21}$--$10^{22}$\,cm$^{-2}$) might not be fully excluded by this cut. However, the selection of optically blue, UV-bright, broad-line type-1 sources independently disfavours significant X-ray obscuration \citep[e.g.,][]{Merloni14}. A spectroscopic analysis of 90 randomly selected objects confirms that the photometric photon indices are unbiased and none of the randomly selected spectra requires intrinsic absorption (Appendix~\ref{app:spec}). Given the aim of this work, it is crucial to expand as much as possible the $\Gamma$ dynamic range while minimising the possible role of obscuration. In Appendix~\ref{app:gamma} we show the effect of the more conservative cut $\Gamma-\Delta\Gamma>1.8$: the single-epoch results are unchanged, but the statistic is reduced. We therefore keep the cut at 1.5, which is the conservative choice for the slope.

After all the cuts described above, the sample contains 7835 quasars (11\,455 \xmm{} detections),  which have a single-epoch $M_{\rm BH}$ measurement from the broad H$\beta$ line (at $z<0.7$; 834 objects), the Mg\,{\sc ii} line (at $0.7<z<2$; 5495), or the C\,{\sc iv} line (at $z>2$; 1506). The different mass subsamples are analysed separately throughout Section \ref{sec:wholesample}, to avoid mixing mass estimators with different calibrations (Sect.~\ref{sec:mbh}). Figure~\ref{fig:gamma_z} shows the distribution of $\Gamma$ as a function of redshift.

\begin{table}[!tbp]
\centering
\caption{Number of objects ($N_{\rm obj}$) and of \xmm{} detections ($N_{\rm det}$) after each selection step.}
\label{tab:cutflow}
\begin{tabular}{lrr}
\hline\hline
Step & $N_{\rm obj}$ & $N_{\rm det}$ \\
\hline
Parent sample & 18\,252 & 28\,756 \\
$\Gamma$ selection & 9877 & 14\,388 \\
Colour selection & 7835 & 11\,455 \\
\hline
\quad H$\beta$, $z<0.7$ & 834 & \\
\quad Mg\,{\sc ii}, $0.7<z<2$ & 5495 & \\
\quad C\,{\sc iv}, $z>2$ & 1506 & \\
\hline
$\geq2$ detections & 1701 & \\
Variability cut & 893 & \\
\hline
\end{tabular}
\end{table}

\section{Eddington-ratio estimates}
\label{sec:eddratio}

The Eddington ratio of an accreting SMBH is
\begin{equation}
  \lambda_{\rm Edd} \equiv \frac{L_{\rm bol}}{L_{\rm Edd}}
  = \frac{L_{\rm bol}}{1.26\times 10^{38}\,(M_{\rm BH}/M_\odot)\,\rm erg\,s^{-1}},
  \label{eq:lambda_edd_def}
\end{equation}
so that $\log\lambda_{\rm Edd} = \log L_{\rm bol} - \log M_{\rm BH} - 38.1$. Both $L_{\rm bol}$ and $M_{\rm BH}$ are derived quantities with prescription-dependent uncertainties. Here we describe different prescriptions that are used throughout the text. 

\subsection{Bolometric corrections}
\label{sec:eddratio_methods}

Many bolometric corrections are available in the literature, derived from different samples and anchored at different wavelengths. Rather than adopting one, we compute $L_{\rm bol}$ with several of them and compare the results. The prescriptions are listed in Table~\ref{tab:kbol} and described in detail in Appendix~\ref{app:kbol}. The optical/UV corrections, $L_{\rm bol}=K\,\nu L_\nu(\lambda)$, fall into two groups. Constant corrections, at 2500\,\AA\ \citep[K13;][]{Krawczyk13}, 4400\,\AA\ \citep[D20;][]{Duras20}, and 5100\,\AA\ \citep[K13$_{5100}$;][]{Krawczyk13, Saccheo23}, only shift $\log L_{\rm bol}$ by an additive constant. Luminosity-dependent corrections change the dynamic range of $L_{\rm bol}$: the empirical power laws of \citet[][N19]{Netzer19}, and the standard thin-disc prediction \citep[][SS]{Shakura73}, $L_{\rm bol}\propto L_{\nu}^{3/2}M_{\rm BH}^{-1}$, which also depends on the black-hole mass. We characterise each prescription by
\begin{equation}
  \kappa \equiv \frac{{\rm d}\log L_{\rm bol}}{{\rm d}\log \nu L_\nu(2500\,{\rm \AA})}
  \label{eq:kappa_def}
\end{equation}
at fixed $M_{\rm BH}$, which is 1 for the constant corrections, 0.8 for N19, and 3/2 for SS. 

We also use a second route, which starts from the monochromatic X-ray luminosity $L_{\rm X}\equiv\nu L_\nu(2\,{\rm keV})$ and derives $L_{2500}$ from the $L_{\rm X}$--$L_{\rm UV}$ relation, $\log L_{\rm X}=\gamma\log L_{2500}+\beta$, with $\gamma=0.591$ and $\beta=8.627$ \citep{Lusso20,Benetti25}; $L_{2500}$ is then converted to $L_{\rm bol}$ with any of the corrections above. We label these prescriptions as $L_{2\,\rm keV}$+K13, +D20, and so on. This route is available for the whole sample at all redshifts. T

This route is available for the whole sample at all redshifts, and does not significantly increase the uncertainty on $L_{\rm bol}$: the observed dispersion of the $L_{\rm X}$--$L_{\rm UV}$ relation is $\approx0.15$\,dex \citep{Risaliti26b}, most of which is due to X-ray variability and inclination rather than to intrinsic differences between quasars \citep{Signorini24}, and it decreases to $\approx0.09$\,dex for luminous quasars with high-quality, dedicated X-ray observations \citep{Sacchi22}. This is smaller than, or comparable to, the scatter of the bolometric corrections themselves (0.2--0.3\,dex; Appendix~\ref{app:kbol}). It is essential for the multi-epoch analysis of Sect.~\ref{sec:multiepoch}, where the optical continuum is not available at the epochs of the X-ray observations. Bolometric corrections based on the integrated 2--10\,keV luminosity, which depends explicitly on $\Gamma$, are not used in the main analysis; we show in Appendix~\ref{app:mock_l210} that they can bias the $\Gamma$--luminosity slope low because of circularity, given that the band luminosity is obtained using the $\Gamma$ itself.

\begin{table}[!tbp]
\centering
\caption{Bolometric corrections used in this work (details in Appendix~\ref{app:kbol}). $\kappa={\rm d}\log L_{\rm bol}/{\rm d}\log L_{2500}$ at fixed $M_{\rm BH}$. For Section \ref{sec:multiepoch}, each correction is also applied to the $L_{2500}$ derived from $L_{\rm  2 keV}$.}
\label{tab:kbol}
\setlength{\tabcolsep}{3.5pt}
\begin{tabular}{llllc}
\hline\hline
Label & Anchor & Form & $\kappa$ \\
\hline
K13 & 2500\,\AA & $K=2.75$ & 1 \\
D20 & 4400\,\AA & $K=5.18$ & 1 \\
K13$_{5100}$ & 5100\,\AA & $K=4.33$ & 1 \\
N19 & 3000\,\AA & $K\propto L^{-0.2}$ &  0.8 \\
SS & 1350--5100\,\AA & $\propto L^{3/2}M_{\rm BH}^{-1}$ &  3/2 \\
\hline
\end{tabular}
\end{table}

\subsection{Black-hole masses}
\label{sec:mbh}
We use the single-epoch virial masses tabulated in DR16Q \citep{Wu22}, $\log(M_{\rm BH}/M_\odot)=A+B\log(L/10^{44}\,{\rm erg\,s^{-1}})+2\log({\rm FWHM}/{\rm km\,s^{-1}})$, with $(A,B)=(0.910,0.50)$ for H$\beta$ and $L_{5100}$ and $(0.660,0.53)$ for C\,{\sc iv} and $L_{1350}$ \citep{Vestergaard06}, and $(0.740,0.62)$ for Mg\,{\sc ii} and $L_{3000}$ \citep{Shen11}. These calibrations form a ladder: the H$\beta$ relation is calibrated on reverberation-mapped AGN, and the Mg\,{\sc ii} and C\,{\sc iv} relations on H$\beta$ or reverberation masses. Each step adds its own uncertainty and possible offsets \citep[e.g.,][]{Shen12, Trakhtenbrot12}, so we never mix mass estimators in a single fit. The formal uncertainties in DR16Q (median 0.08--0.09\,dex) reflect only the precision of the line and continuum measurements, and do not include the systematic uncertainty of the virial calibration, $\sim$\,0.4\,dex \citep{Vestergaard06, Shen12}.

\section{Population analysis}
\label{sec:wholesample}

\begin{figure*}[!tp]
\centering
\includegraphics[width=0.95\linewidth,clip]{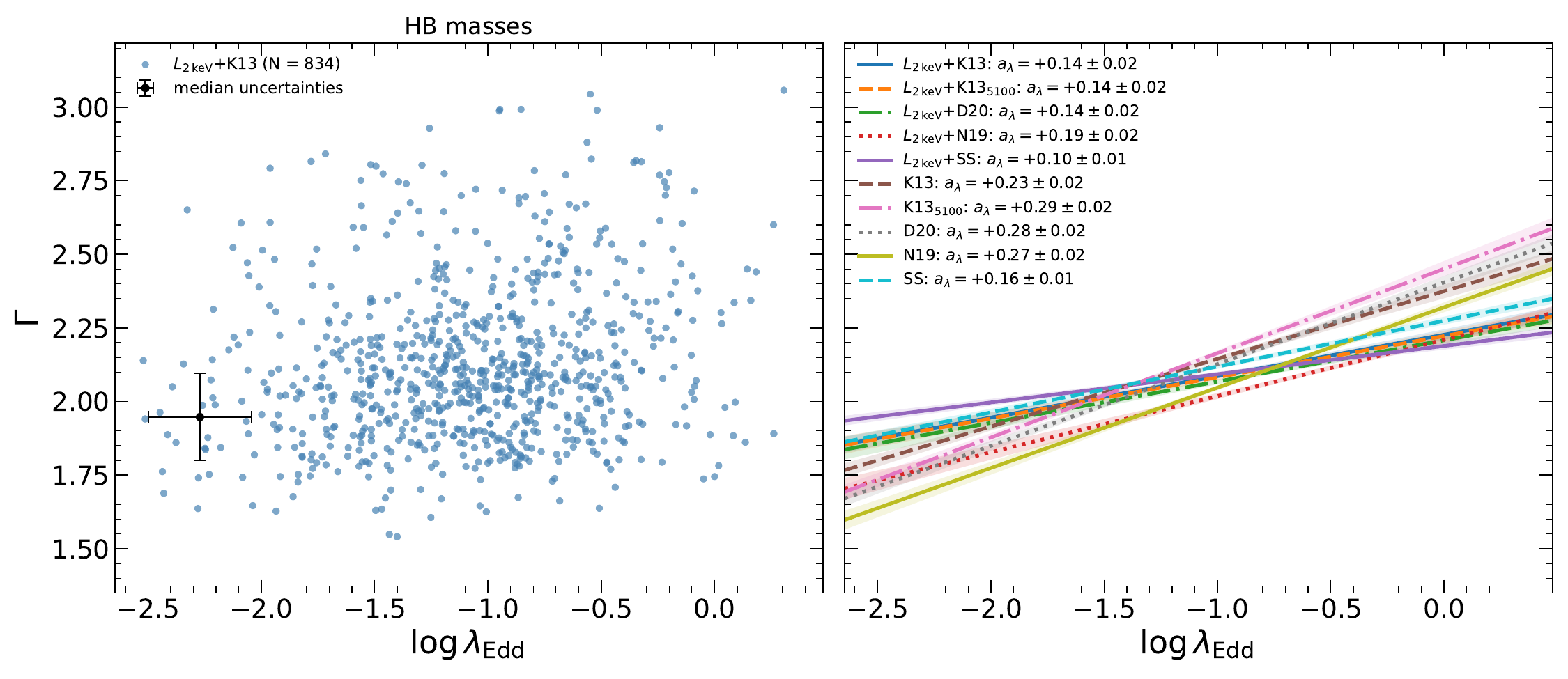}
\includegraphics[width=0.95\linewidth,clip]{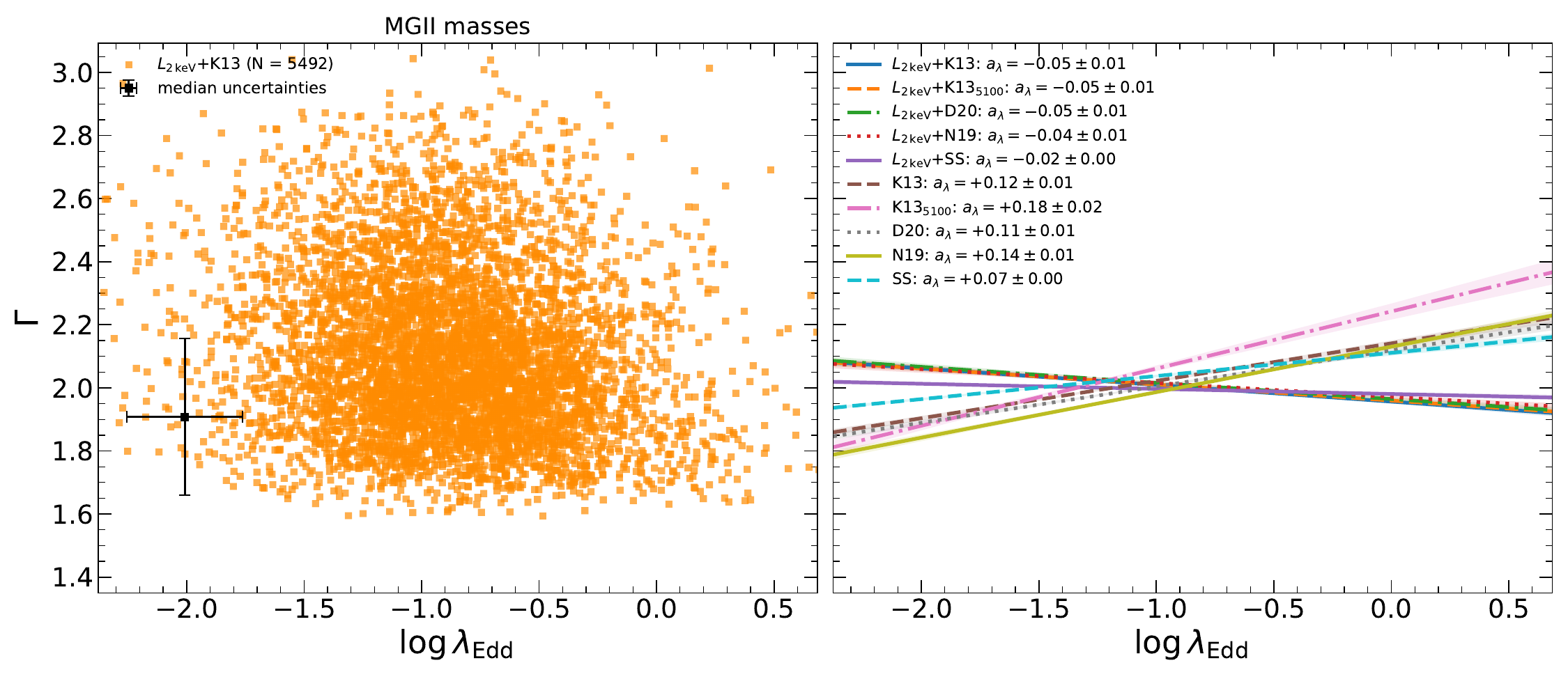}
\includegraphics[width=0.95\linewidth,clip]{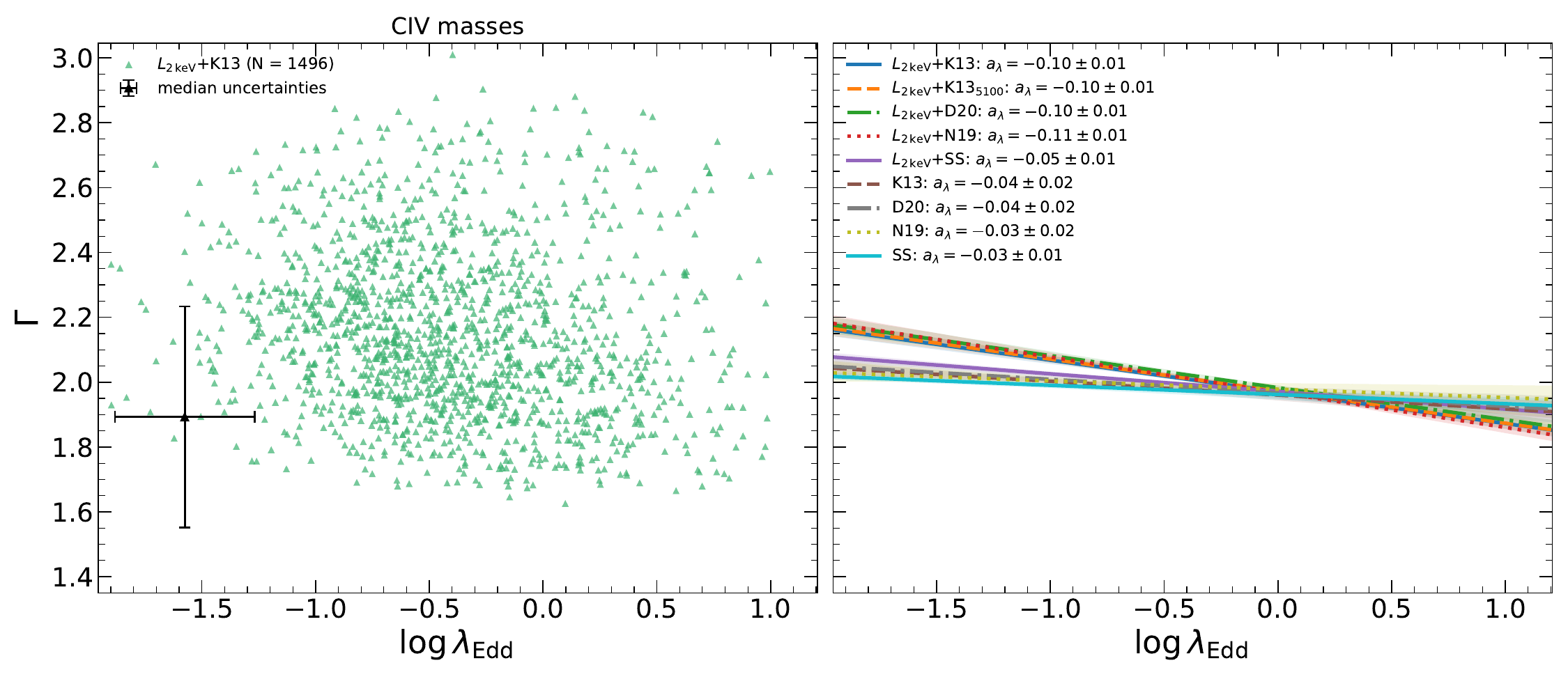}

\caption{Whole-sample $\Gamma$--$\log\lambda_{\rm Edd}$ fits for the H$\beta$ (top), Mg\,{\sc ii} (middle), and C\,{\sc iv} (bottom) subsamples. Left: data, with $\lambda_{\rm Edd}$ computed with the $L_{2\,\rm keV}$+K13 prescription; the cross shows the median uncertainties. Right: best-fit relations for the prescriptions of Table~\ref{tab:kbol}, with 68\% posterior bands.}
\label{fig:wholesample_Hb_MgII_CIV}
\end{figure*}

Our first approach to measuring the $\Gamma$--$\lambda_{\rm Edd}$ relation is a single regression of $\Gamma$ against $\log\lambda_{\rm Edd}$ across the entire cleaned sample, as in most previous studies. We compute $\log\lambda_{\rm Edd}$ with each of the bolometric corrections of Sect.~\ref{sec:eddratio_methods} in turn, and separately for each of the three SE mass estimators of Sect.~\ref{sec:mbh}. With respect to existing studies, which typically include a few hundred AGN at most \citep[e.g.,][]{Shemmer06, Shemmer08, Kelly08, Risaliti09, Brightman13, Fanali13, Trakhtenbrot17}, our sample is more than an order of magnitude larger and homogeneously selected. We model the relation as $\Gamma = a_{\lambda}\,\log\lambda_{\rm Edd} + b_\lambda$, with an additional intrinsic scatter $\sigma_{\rm int}$ in the $\Gamma$ direction, and use the effective-variance likelihood. Details are given in Appendix \ref{app:fitting}.

\subsection{Results}
\label{sec:wholesample_results}
Table~\ref{tab:wholesample_summary} reports the slope, intercept, and intrinsic scatter for each bolometric correction and each mass subsample, together with the observed dispersion of $\log\lambda_{\rm Edd}$, $\sigma_{\rm obs}$; the fits are shown in Fig.~\ref{fig:wholesample_Hb_MgII_CIV}. The intrinsic scatter is significantly different from zero in all cases.\\
Not all prescriptions are available over the full redshift range of each subsample, because the anchor wavelength must fall within the SDSS spectrum: K13$_{5100}$ includes 1196 of the 5495 Mg\,{\sc ii} objects and none of the C\,{\sc iv} ones, while D20 and N19 include about 890 of the 1506 C\,{\sc iv} objects. The monochromatic X-ray route is available for the whole sample.\\
The slopes depend on both the mass indicator and the bolometric correction. For H$\beta$ masses all prescriptions give positive slopes, $a_\lambda=0.10$--0.29 (0.10--0.19 for the X-ray routes). For Mg\,{\sc ii} masses, the X-ray routes give slopes close to zero or slightly negative ($a_\lambda=-0.05$ to $-0.01$), while the optical/UV corrections give $a_\lambda=0.07$--0.18. For C\,{\sc iv} masses, all slopes are negative ($a_\lambda=-0.11$ to $-0.03$). Within each subsample, the slope decreases as $\sigma_{\rm obs}$ increases: the SS prescriptions, in which the mass enters $\lambda_{\rm Edd}$ twice, have the largest $\sigma_{\rm obs}$ (0.65--0.84\,dex) and the flattest slopes among the optical/UV routes. The C\,{\sc iv} results should be treated with particular care, since C\,{\sc iv}-based virial masses carry well-known systematics \citep{Shen11, Trakhtenbrot12}; we keep this subsample because it is the only one covering $z>2$. 

\begin{figure*}[!tp]
\centering
\includegraphics[width=\linewidth,clip]{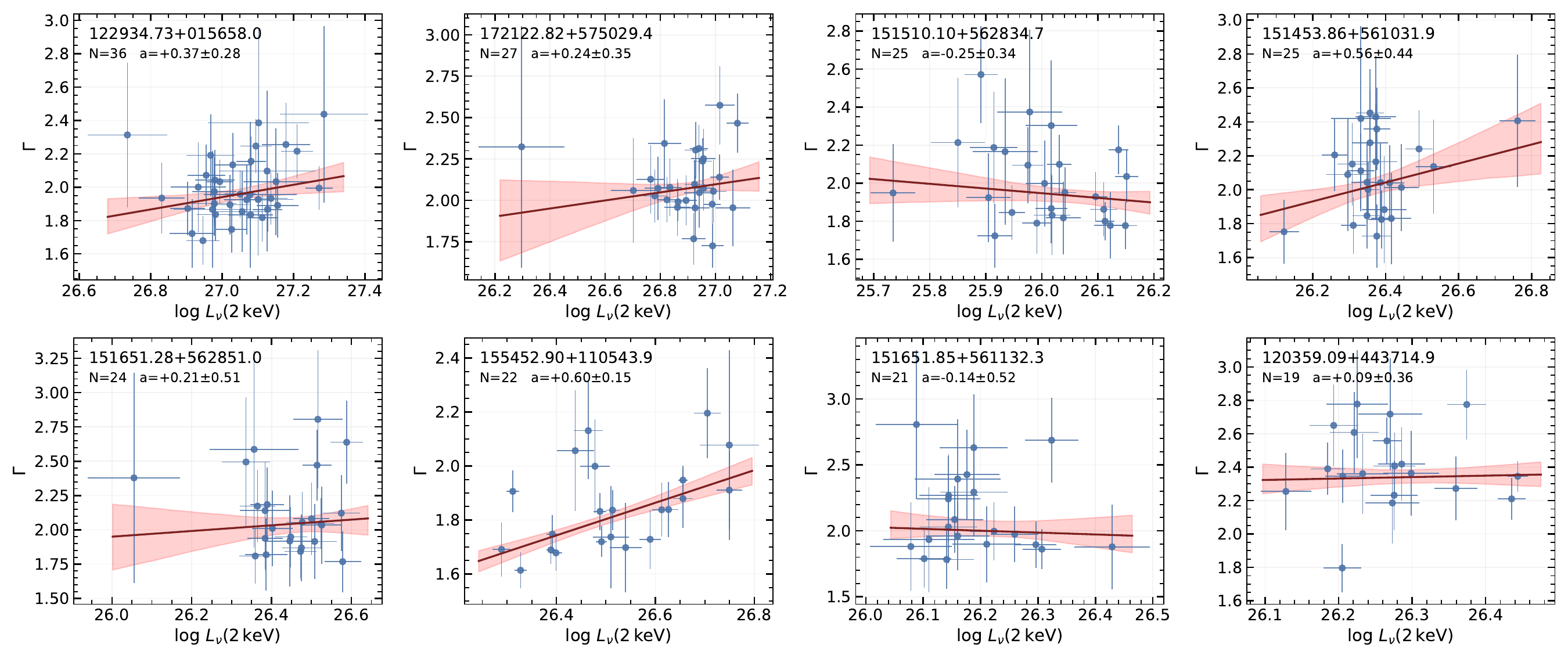}
\caption{$\Gamma$--$\log L_{\rm X}$ fits for the eight objects with the most \xmm~ observations, illustrating the object-to-object diversity in the recovered slopes $a_X$.}
\label{fig:multiple_8}
\end{figure*}

\begin{table*}[!tp]
\centering
\caption{Whole-sample effective-variance fit results. Slopes and intercepts are posterior medians; uncertainties correspond to the 16th--84th percentiles. $\sigma_{\rm int}$ is the additional parameter in the fit required in addition to uncertainty to explain the total observed scatter.  $\sigma_{\rm obs}$ is the observed scatter of $\log\lambda_{\rm Edd}$ in the subsample. Labels as in Table~\ref{tab:kbol}; $L_{2\,\rm keV}$+ denotes the monochromatic X-ray route.}
\label{tab:wholesample_summary}
\setlength{\tabcolsep}{3.5pt}
\begin{tabular}{lccccc}
\hline\hline
Method & $N$ & $a_{\lambda}$ & $b_\lambda$ & $\sigma_{\rm int}$ & $\sigma_{\rm obs}$ \\
\hline
\multicolumn{6}{c}{H$\beta$ subsample ($z<0.7$, $N_{\rm tot}=834$)} \\
\hline
$L_{2\,\rm keV}$+K13            &   834 & $+0.141\pm0.015$ & $2.227\pm0.018$ & $0.187\pm0.007$ & $0.535$ \\
$L_{2\,\rm keV}$+K13$_{5100}$   &   834 & $+0.140\pm0.016$ & $2.222\pm0.018$ & $0.186\pm0.007$ & $0.535$ \\
$L_{2\,\rm keV}$+D20            &   834 & $+0.140\pm0.016$ & $2.208\pm0.016$ & $0.187\pm0.007$ & $0.535$ \\
$L_{2\,\rm keV}$+N19            &   834 & $+0.191\pm0.018$ & $2.210\pm0.014$ & $0.180\pm0.007$ & $0.466$ \\
$L_{2\,\rm keV}$+SS             &   799 & $+0.099\pm0.009$ & $2.126\pm0.010$ & $0.164\pm0.007$ & $0.840$ \\
K13              &   806 & $0.230\pm0.017$ & $2.375\pm0.023$ & $0.171\pm0.007$ & $0.479$ \\
K13$_{5100}$     &   834 & $0.286\pm0.018$ & $2.451\pm0.025$ & $0.164\pm0.007$ & $0.414$ \\
D20              &   834 & $0.277\pm0.018$ & $2.405\pm0.022$ & $0.165\pm0.006$ & $0.425$ \\
N19              &   806 & $0.273\pm0.018$ & $2.320\pm0.018$ & $0.164\pm0.007$ & $0.434$ \\
SS               &   832 & $0.156\pm0.010$ & $2.182\pm0.010$ & $0.160\pm0.006$ & $0.792$ \\
\hline
\multicolumn{6}{c}{Mg\,{\sc ii} subsample ($0.7<z<2$, $N_{\rm tot}=5495$)} \\
\hline
$L_{2\,\rm keV}$+K13            &  5492 & $-0.053\pm0.007$ & $1.956\pm0.006$ & $0.125\pm0.003$ & $0.468$ \\
$L_{2\,\rm keV}$+K13$_{5100}$   &  5490 & $-0.051\pm0.007$ & $1.960\pm0.006$ & $0.125\pm0.003$ & $0.467$ \\
$L_{2\,\rm keV}$+D20            &  5490 & $-0.051\pm0.007$ & $1.965\pm0.006$ & $0.125\pm0.003$ & $0.467$ \\
$L_{2\,\rm keV}$+N19            &  5492 & $-0.044\pm0.008$ & $1.972\pm0.006$ & $0.127\pm0.003$ & $0.395$ \\
$L_{2\,\rm keV}$+SS             &  5343 & $-0.008\pm0.005$ & $1.996\pm0.004$ & $0.125\pm0.003$ & $0.698$ \\
K13              &  5495 & $0.119\pm0.008$ & $2.141\pm0.010$ & $0.119\pm0.003$ & $0.378$ \\
K13$_{5100}$     &  1196 & $0.181\pm0.018$ & $2.242\pm0.025$ & $0.128\pm0.006$ & $0.363$ \\
D20              &  5495 & $0.115\pm0.008$ & $2.119\pm0.010$ & $0.120\pm0.003$ & $0.371$ \\
N19              &  5495 & $0.144\pm0.010$ & $2.131\pm0.010$ & $0.118\pm0.003$ & $0.329$ \\
SS               &  5489 & $0.074\pm0.005$ & $2.067\pm0.006$ & $0.118\pm0.003$ & $0.653$ \\
\hline
\multicolumn{6}{c}{C\,{\sc iv} subsample ($z>2$, $N_{\rm tot}=1506$)} \\
\hline
$L_{2\,\rm keV}$+K13            &  1496 & $-0.098\pm0.013$ & $1.970\pm0.007$ & $0.085\pm0.003$ & $0.532$ \\
$L_{2\,\rm keV}$+K13$_{5100}$   &  1492 & $-0.099\pm0.012$ & $1.972\pm0.008$ & $0.085\pm0.003$ & $0.528$ \\
$L_{2\,\rm keV}$+D20            &  1485 & $-0.099\pm0.013$ & $1.983\pm0.007$ & $0.086\pm0.003$ & $0.521$ \\
$L_{2\,\rm keV}$+N19            &  1498 & $-0.109\pm0.013$ & $1.969\pm0.007$ & $0.086\pm0.004$ & $0.484$ \\
$L_{2\,\rm keV}$+SS             &  1224 & $-0.057\pm0.011$ & $2.006\pm0.008$ & $0.086\pm0.004$ & $0.690$ \\
K13              &  1391 & $-0.043\pm0.017$ & $1.960\pm0.017$ & $0.095\pm0.007$ & $0.380$ \\
D20              &   891 & $-0.042\pm0.020$ & $1.967\pm0.018$ & $0.090\pm0.007$ & $0.417$ \\
N19              &   894 & $-0.026\pm0.023$ & $1.978\pm0.019$ & $0.091\pm0.008$ & $0.406$ \\
SS               &  1478 & $-0.027\pm0.010$ & $1.980\pm0.009$ & $0.094\pm0.007$ & $0.678$ \\
\hline
\end{tabular}
\end{table*}

Contrary to the naive expectation that a larger sample would yield a better-defined relation, the recovered slopes are flatter than, or at most comparable to, the canonical $a_\lambda\approx0.3$, and they are not mutually consistent: their sign depends on the mass estimator and on the chosen bolometric correction. In Appendix~\ref{app:gamma} we show that a stricter cut on the photon index, $\Gamma-\Delta\Gamma>1.8$, gives analogous results.

\subsection{Mass noise and regression dilution}
\label{sec:wholesample_MBHscatter}
These results can be understood by considering the composition of the $\log\lambda_{\rm Edd}$ axis. \cite{Risaliti26} argued that the intrinsic dispersion of $\log\lambda_{\rm Edd}$ in optically selected SDSS quasars is narrow, $\lesssim$\,0.15\,dex, based on the small dispersion of the broad-line Baldwin effect and on the narrow distribution of disc temperatures inferred from stacked UV spectra. Our sample is a subsample of theirs, and we assume a similar intrinsic spread. The observed spread is instead 0.33--0.54\,dex for the empirical corrections (Table~\ref{tab:wholesample_summary}), comparable to the systematic uncertainty of SE virial masses, $\sim$\,0.4\,dex \citep{Vestergaard06, Shen12}, which is not included in the formal errors. Most of the variance of the $x$ axis is therefore mass noise. Errors in the independent variable attenuate a regression slope by the factor $\sigma^2_{\rm true}/(\sigma^2_{\rm true}+\sigma^2_{\rm noise})$; with $\sigma_{\rm true}\approx0.15$\,dex and $\sigma_{\rm obs}\approx0.4$--0.5\,dex this factor is $\approx0.1$, so that an intrinsic slope of 0.2, as measured in Sect.~\ref{sec:multiepoch}, would appear as $a_\lambda\approx0.02$.

We tested this with forward-modelled mock samples that reproduce the redshift distribution, the effective X-ray and optical flux limits, the $\Gamma$-dependent selection and the $\Gamma$ uncertainties of each subsample, with an intrinsic $\lambda_{\rm Edd}$ spread of 0.15\,dex and virial-mass noise of 0.4\,dex (Appendix~\ref{app:mock}). For injected slopes between 0 and 0.4, the recovered slopes are between $-0.02$ and $+0.03$ in all three subsamples and for both the K13 and the $L_{2\,\rm keV}$+K13 routes. Two conclusions follow. First, with realistic mass noise a whole-sample regression cannot recover a $\Gamma$--$\lambda_{\rm Edd}$ slope of the size measured in Sect.~\ref{sec:multiepoch}. Second, neither the flux limits nor the $\Gamma$ selection can produce the differences between the subsamples: selection alone yields slopes consistent with zero, with a small negative offset ($\approx-0.02$) from the $\Gamma$ cut, and cannot produce the positive slopes of the H$\beta$ subsample or the negative ones of the C\,{\sc iv} subsample.

Since independent noise can only flatten a slope, the non-zero slopes of Table~\ref{tab:wholesample_summary} require errors on the $\log\lambda_{\rm Edd}$ axis that are correlated with $\Gamma$. The line width is the natural candidate, as it enters the virial mass as $2\log{\rm FWHM}$. For H$\beta$ and Mg\,{\sc ii}, $\Gamma$ is known to correlate with the line width: R09 found a $\Gamma$--FWHM correlation almost as strong as the $\Gamma$--$\lambda_{\rm Edd}$ one, and the same correlation strength for any combination ${\rm FWHM}^{-2}L^{\beta}$ with $0<\beta<0.8$, concluding that the luminosity term does not contribute to their correlation. A dependence of $\Gamma$ on FWHM that is not purely mediated by the true $\lambda_{\rm Edd}$ (for example through the geometry or orientation of the BLR) makes the mass errors correlated with $\Gamma$ and produces a positive population slope even when the mass noise is large. 
For C\,{\sc iv}, the line width is affected by outflows whose strength increases with $\lambda_{\rm Edd}$, as traced by the C\,{\sc iv} blueshift \citep[e.g.,][]{Coatman17}: highly accreting sources have broader lines, overestimated masses and underestimated $\lambda_{\rm Edd}$, which drives the population slope negative. A second effect may contribute to the positive slopes of the H$\beta$ subsample. At $z<0.7$ the 1--4.5\,keV band used to measure $\Gamma$ corresponds to rest-frame energies below $\approx$\,2--3\,keV, where the soft excess can steepen the photometric photon index. The strength of the soft excess increases with $\lambda_{\rm Edd}$, while it does not correlate with $M_{\rm BH}$ or with the primary X-ray luminosity \citep[e.g.,][]{Boissay16, Chen25}, so that its contribution to $\Gamma$ may depend on the accretion parameters differently from that of the primary power law. 
In either case, the population slopes are dominated by effects specific to each subsample, through the virial masses or the spectral band sampled, rather than by the $\Gamma$--$\lambda_{\rm Edd}$ relation, which explains why their sign depends on the emission line used.\\
Overall, we conclude that a whole-sample regression with SE virial masses cannot measure the slope of the $\Gamma$--$\log\lambda_{\rm Edd}$ relation, regardless of the bolometric correction: the result is diluted by mass noise and dominated by the systematics of the mass estimator. The way around this limitation is to remove the black-hole mass from the $x$-axis.

\section{Per-object analysis}
\label{sec:multiepoch}

For a single source observed at multiple epochs, $M_{\rm BH}$ is fixed and enters only the intercept of the $\Gamma$--$\log\lambda_{\rm Edd}$ relation. Changes in continuum luminosity then trace changes in the accretion state of that object, without the scatter introduced by SE mass measurements. 
Here we implement this approach by measuring the $\Gamma$--$\log L_{\rm X}$ relation object by object and translating the result to $\lambda_{\rm Edd}$. Throughout this section we distinguish between the accretion rate $\dot M$ and the Eddington ratio: at fixed mass the luminosity tracks $\dot M$, but whether $L_{\rm bol}$, and hence $\lambda_{\rm Edd}$, follows it one-to-one is an assumption (Sect.~\ref{sec:conversion}).

We start from the 1701 objects in our sample with at least two \xmm{} observations. AGN variability is stochastic \citep{Uttley05, Sartori18, Paolillo25} and the multiple observations in the serendipitous catalogue are randomly spaced in time, so only objects whose $\log L_{\rm X}$ varies appreciably across epochs can constrain a per-object slope. We therefore retain objects with ${\rm std}(\log L_{{\rm X},k}) / {\rm median}(\sigma_{{\rm X},k}) > 1$, i.e.\ whose epoch-to-epoch scatter exceeds the typical per-epoch uncertainty on $\log L_{\rm X}$. This selection reduces the sample to 893 objects. Because it favours the most variable sources, it also changes the composition of the sample: the selected objects have lower redshift (median $z=1.08$ against 1.59 for the rejected ones), lower X-ray luminosity, more epochs, and slightly lower black-hole mass and Eddington ratio (Appendix~\ref{app:varcut}). The mean slope, however, depends only weakly on the threshold: it ranges from $0.33\pm0.03$ with no variability cut to $0.36\pm0.03$ for a threshold of 2 (Table~\ref{tab:varcut}). A mild increase is expected, since objects whose luminosity changes are dominated by noise dilute the slope.

\subsection{Per-object $\Gamma$--$\log L_{\rm X}$ fits}
\label{sec:perobject}

Our strategy consists of three steps: 
{\it (i)} fit the relation $\Gamma=a_{X,j}\log L_{\rm X}+b_{X,j}$ for each object $j$;
{\it (ii)} combine the per-object posteriors on $a_{X,j}$ into the population mean $\langle a_X\rangle$ and intrinsic dispersion;
{\it (iii)} translate $\langle a_X\rangle$ into a $\Gamma$--$\log\lambda_{\rm Edd}$ slope.\\
Several caveats apply. First, the number of observations per object is limited: of the 893 objects, 462 have at least 3 observations, 164 at least 5, and only 45 at least 10, so most per-object posteriors are broad. Second, the time sampling is heterogeneous, with intervals between observations ranging from days to over a decade, and the same baseline samples very different parts of the variability power spectrum for a $10^7$ or a $10^{9.5}\,M_\odot$ black hole \citep[e.g.,][]{Vaughan03, Vagnetti11, Lanzuisi14}. Third, any physical relation is expected to have intrinsic scatter. As a result the per-object slopes vary substantially from object to object.
For each object we fit the relation with the same effective-variance likelihood as described in Appendix \ref{app:fitting} and as used in the previous Section, without the intrinsic-scatter term, since the measurement uncertainties are sufficient to explain the observed scatter around individual relations. \\
Figure~\ref{fig:multiple_8} shows eight objects with at least ten observations each. Most of them show a positive slope, but the slopes cover a broad range and several are consistent with zero. 

\subsection{Combining the per-object posteriors}
\label{sec:stacking}

\begin{figure}[!htbp]
\centering
\includegraphics[width=0.9\linewidth,clip]{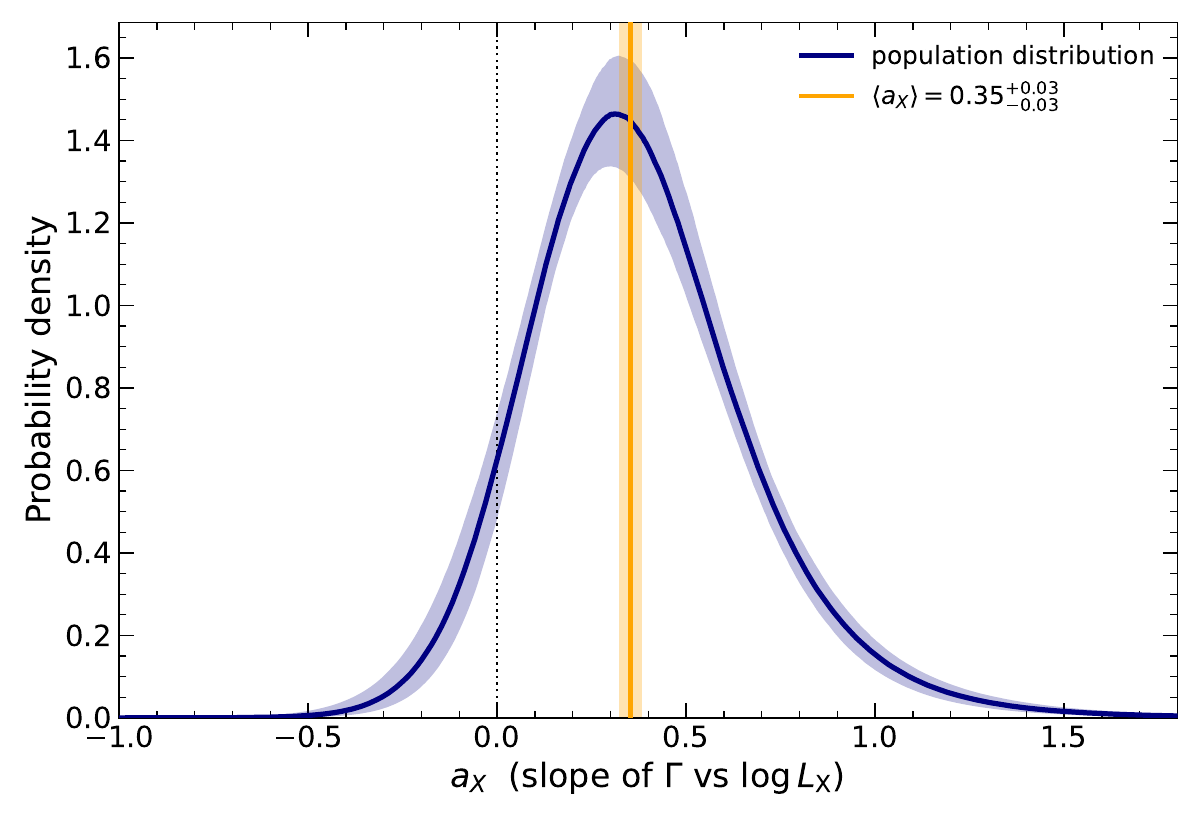}
\caption{Distribution of the per-object slopes $a_{X,j}$ inferred with \texttt{PosteriorStacker} for the 893 objects with at least two observations (blue: median and 68\% band of the population distribution), and population-mean slope $\langle a_X\rangle$ with its 68\% uncertainty (orange). Grey curves show the posteriors of a random subset of individual objects, rescaled for display.}
\label{fig:posteriorstacker}
\end{figure}

We combine the per-object posteriors into a population distribution of slopes with the public code \texttt{PosteriorStacker} \citep{Baronchelli20}. The code fits a parent distribution to the full set of posteriors, marginalising over the measurement uncertainty encoded in each of them without assuming that they are Gaussian, so that precise per-object measurements carry more weight and uninformative posteriors are automatically down-weighted. We model the parent distribution as a Gaussian, whose mean gives the population-mean slope $\langle a_X\rangle$ and whose width gives the intrinsic object-to-object dispersion of the slopes, $\sigma_a$; further details are given in Appendix~\ref{app:fitting}. \\
Table~\ref{tab:multiepoch_lx_alx} reports the results. For the full sample we find $\langle a_X\rangle=0.35\pm0.03$ (Fig.~\ref{fig:posteriorstacker}). The result is stable against the minimum number of epochs, with $0.33\pm0.04$ for the 164 objects with $N_{\rm min}=5$ and $0.34\pm0.08$ for the 45 with $N_{\rm min}=10$, while the uncertainty on the mean grows as the sample shrinks. It is also stable against the variability selection (Sect.~\ref{sec:multiepoch}). In quartiles of black-hole mass the slope shows no monotonic trend, while it decreases mildly with X-ray luminosity (Appendix~\ref{app:varcut}).\\
The dispersion of the per-object slopes, $\sigma_a\approx0.28$, is about 80\% of the mean: individual quasars display different $\Gamma$--$L_{\rm X}$ couplings, some consistent with no response at all. Part of this diversity is due to measurement effects (few epochs, photometric $\Gamma$, heterogeneous time sampling), and part may be physical, for example different coronal geometries \citep[e.g.,][]{Laurenti24}.

\begin{table}[!tbp]
\centering
\caption{Multi-epoch population slopes in the $\Gamma$--$\log L_{\rm X}$ plane. $N_{\rm min}$ is the minimum number of observations and $N$ the number of objects passing the variability cut; $\langle a_X\rangle$ is shown with uncertainty from the 16th--84th percentiles; $\sigma_a$ is the intrinsic dispersion of the population.}
\label{tab:multiepoch_lx_alx}
\begin{tabular}{rccc}
\hline\hline
$N_{\rm min}$ & $N$ & $\langle a_X\rangle$ &  $\sigma_a$ \\
\hline
 2 &  893 & $0.35\pm0.03$ &  $0.28$ \\
 3 &  462 & $0.34\pm0.03$ &  $0.25$ \\
 5 &  164 & $0.33\pm0.04$ &  $0.24$ \\
10 &   45 & $0.34\pm0.08$ &  $0.33$ \\
\hline
\end{tabular}
\end{table}

\subsection{Conversion to $\Gamma$--$\log\lambda_{\rm Edd}$}
\label{sec:conversion}

The conversion is applied to the population mean $\langle a_X\rangle$, and involves two steps. First, the $L_{\rm X}$--$L_{\rm UV}$ relation maps a change in $\log L_{\rm X}$ into a change $\Delta\log L_{2500}=\Delta\log L_{\rm X}/\gamma$, so that ${\rm d}\Gamma/{\rm d}\log L_{2500}=\langle a_X\rangle\,\gamma$. Second, the bolometric correction maps $\Delta\log L_{2500}$ into $\Delta\log L_{\rm bol}=\kappa\,\Delta\log L_{2500}$ (Eq.~\ref{eq:kappa_def}), and at fixed mass $\Delta\log\lambda_{\rm Edd}=\Delta\log L_{\rm bol}$. Hence
\begin{equation}
  \langle a_\lambda\rangle = \langle a_X\rangle\,\frac{\gamma}{\kappa}\,.
  \label{eq:slope_conversion}
\end{equation}
A constant bolometric correction only shifts the intercept, so K13, D20 and K13$_{5100}$ give identical slopes by construction; the conversion from 2500\,\AA\ to another anchor wavelength with a fixed UV slope is also a constant offset. Since the conversion is linear, the relative width of the slope distribution is preserved.\\
Table~\ref{tab:multiepoch_lx_aedd} reports the results. For our reference prescription (K13, $\kappa=1$), the full sample gives $\langle a_\lambda\rangle=0.21\pm0.02$, independently of $N_{\rm min}$. This is flatter than the slopes of S08, R09 and B13 (0.31--0.32), close to those of \cite{Liu21} ($0.27\pm0.04$) and \cite{Laurenti24} ($0.23\pm0.04$), and steeper than those of T17 ($0.17\pm0.04$) and \cite{Trefoloni23} ($0.16\pm0.03$). T17, derive $L_{\rm bol}$ from the 2--10\,keV luminosity, which might be why the detected slope is lower than in other results (Appendix~\ref{app:mock_l210}).
Equation~\ref{eq:slope_conversion} assumes that the population values of $\gamma$ and $\kappa$ also describe the response of an individual object at fixed mass, which is not guaranteed. The slope of the $L_{\rm X}$--$L_{\rm UV}$ relation does not evolve with redshift and does not depend strongly on the choice of X-ray and UV indicators \citep[e.g.,][]{Signorini23}, but these results refer to the quasar population, and do not necessarily apply to the variations of a single source. The only direct measurement of the per-object slope, obtained from six years of simultaneous UV and X-ray monitoring of the Seyfert ESO\,511-G030, gives $\gamma=0.74\pm0.07$ \citep{Middei26}, somewhat steeper than the population value. Allowing $\gamma$ to range conservatively between 0.5 and 0.74 changes $\langle a_\lambda\rangle$ (K13) from 0.18 to 0.26. \\
Luminosity-dependent empirical corrections such as N19 ($\kappa=0.8$--0.9) are not applicable here: their power laws describe how the mean correction changes across a population of objects with different masses and spins, whereas at fixed mass and spin the underlying disc models give a correction that increases with the accretion rate, i.e.\ $\kappa>1$ \citep{Netzer19}. The standard thin disc predicts $\kappa=3/2$, which would lower the slope to $\langle a_\lambda\rangle=0.14\pm0.01$ (SS column of Table~\ref{tab:multiepoch_lx_aedd}). Combining these uncertainties, we adopt $\langle a_\lambda\rangle\approx0.2$ as our reference value, with a plausible range of 0.14--0.26 dominated by the unknown per-object values of $\gamma$ and $\kappa$ rather than by statistics.

\begin{table}[!tbp]
\centering
\caption{Multi-epoch $\Gamma$--$\log\lambda_{\rm Edd}$ slopes $\langle a_\lambda\rangle=\langle a_X\rangle\,\gamma/\kappa$ (Eq.~\ref{eq:slope_conversion}), with $\gamma=0.591$, for the $\langle a_X\rangle$ of Table~\ref{tab:multiepoch_lx_alx}. The constant corrections (K13, D20, K13$_{5100}$) all have $\kappa=1$ and give identical slopes. Uncertainties include the uncertainty on $\gamma$.}
\label{tab:multiepoch_lx_aedd}
\begin{tabular}{rccc}
\hline\hline
$N_{\rm min}$ & $N$ & $\kappa=1$ & SS ($\kappa=3/2$) \\
\hline
 2 &  893 & $0.21\pm0.02$ & $0.14\pm0.01$ \\
 3 &  462 & $0.20\pm0.02$ & $0.13\pm0.01$ \\
 5 &  164 & $0.19\pm0.03$ & $0.13\pm0.02$ \\
10 &   45 & $0.20\pm0.05$ & $0.13\pm0.03$ \\
\hline
\end{tabular}
\end{table}

\subsection{Dependence on the photon index measurements}
\label{sec:snr}

\begin{table}[!tp]
\centering
\caption{Multi-epoch slopes ($N_{\rm min}=2$) for increasing thresholds on the signal-to-noise ratio of the per-epoch photon index, $\Gamma/\sigma_\Gamma$. $\langle a_X\rangle$ uncertainties are half the 16th--84th percentile range; $\langle a_\lambda\rangle=\langle a_X\rangle\,\gamma/\kappa$ as in Table~\ref{tab:multiepoch_lx_aedd}.}
\label{tab:snr}
\setlength{\tabcolsep}{5pt}
\begin{tabular}{lrcccc}
\hline\hline
$\Gamma/\sigma_\Gamma$ & $N$ & $\langle a_X\rangle$ & K13 ($\kappa=1$) &  SS ($\kappa=3/2$) \\
\hline
--     &  893 & $0.35\pm0.03$ & $0.21\pm0.02$ & $0.14\pm0.01$ \\
$>7$   &  797 & $0.37\pm0.03$ & $0.22\pm0.01$ & $0.14\pm0.01$ \\
$>10$  &  528 & $0.43\pm0.04$ & $0.25\pm0.02$ & $0.17\pm0.01$ \\
$>13$  &  336 & $0.44\pm0.04$ & $0.26\pm0.02$ & $0.17\pm0.02$ \\
$>15$  &  256 & $0.43\pm0.04$ & $0.25\pm0.02$ & $0.17\pm0.01$ \\
\hline
\end{tabular}
\end{table}
The photometric photon indices of our serendipitous sources are measured with very different precision. The exposure time, off-axis angle, and background of each observation are set by the programme for which it was taken, not by our targets, and they vary from one epoch to another of the same source; the per-epoch uncertainty on $\Gamma$ has a median of 0.19 and a broad distribution. The likelihood accounts for these uncertainties, but a relation can still be diluted if many measurements are very noisy. We therefore repeated the multi-epoch analysis retaining only epochs whose photon index is measured with a signal-to-noise ratio $\Gamma/\sigma_\Gamma$ above a given threshold, and then applying the same variability cut and stacking as for the fiducial sample. This selection differs in nature from a selection on the epoch-to-epoch spread of $\Gamma$ within an object, which would select on the dependent variable and bias the slope: $\Gamma/\sigma_\Gamma$ measures the quality of an observation rather than the value or the variation of $\Gamma$.

The results are listed in Table~\ref{tab:snr}. The slope increases from $\langle a_X\rangle=0.35\pm0.03$ with no selection to $0.34\pm0.03$ and $0.37\pm0.03$ for thresholds of 5 and 7, and then reaches a plateau at $\langle a_X\rangle\approx0.43$ for thresholds of 10 and above, while the sample decreases from 528 to 256 objects. The corresponding Eddington-ratio slopes are $\langle a_\lambda\rangle\approx0.25$ for K13, and 0.17 for SS.

This selection is not completely free of biases. Epochs with a precise photon index are those with more counts, so the selection favours brighter states within each object and brighter objects in the sample, and therefore correlates slightly with luminosity and redshift. The fact that the slope reaches a stable value over a factor of two in sample size, however, suggests that the plateau reflects the intrinsic response rather than a progressive selection effect. We therefore regard $\langle a_X\rangle=0.35\pm0.03$ as a conservative estimate, and $\langle a_X\rangle\approx0.43$ as the best estimate of the response of $\Gamma$ when it is measured precisely. For the reference conversion (constant bolometric correction and $\gamma=0.591$), the latter corresponds to $\langle a_\lambda\rangle\approx0.25$; the per-object uncertainties on $\gamma$ and $\kappa$ discussed in Sect.~\ref{sec:conversion} widen this to a plausible range of 0.17--0.32.

\subsection{Comparison with the whole-sample analysis}
\label{sec:multiepoch_comparison}

\begin{figure*}[!htbp]
\centering
\includegraphics[width=0.6\linewidth,clip]{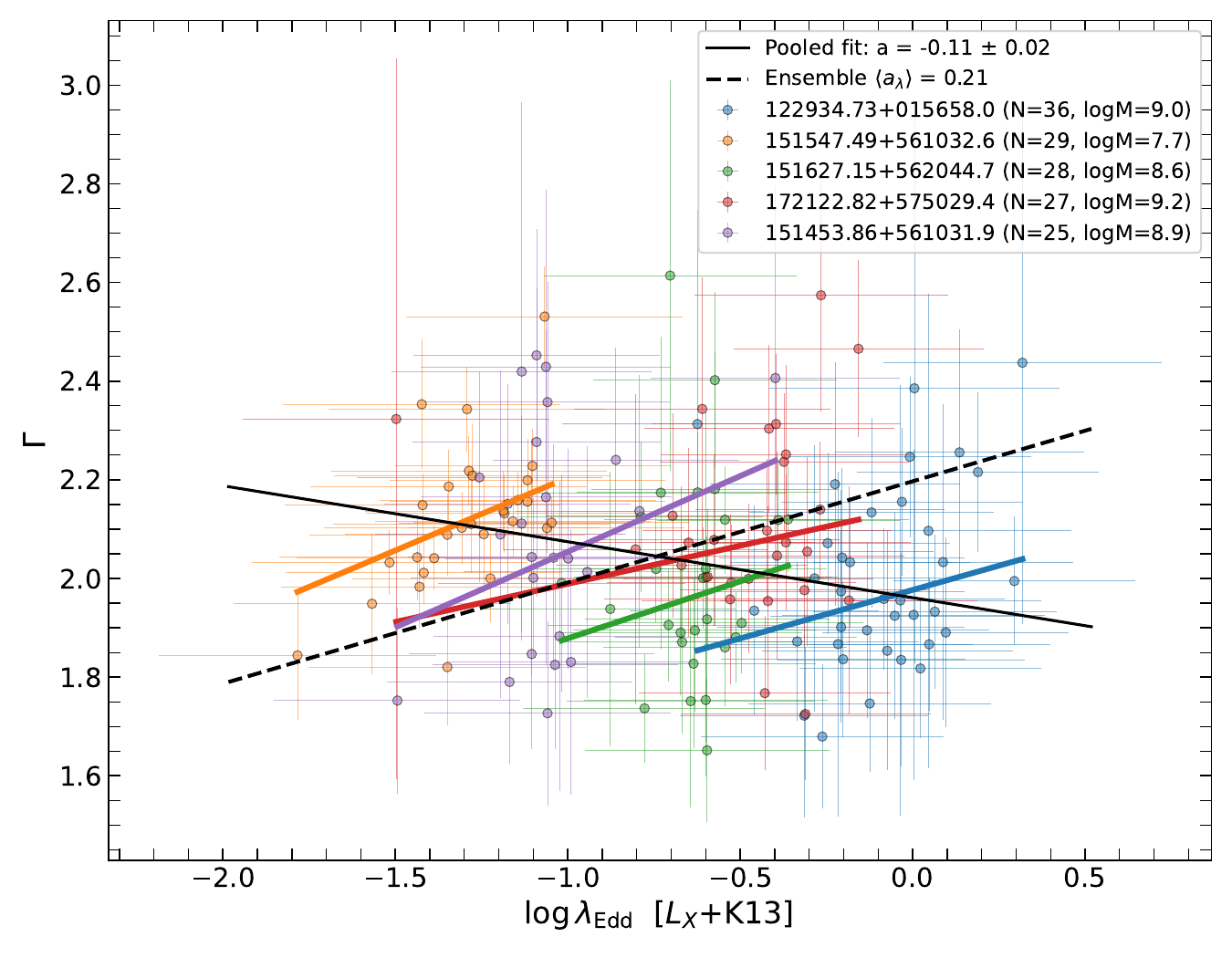}
\caption{$\Gamma$ vs.\ $\log\lambda_{\rm Edd}$ for all epochs of the five objects with the largest number of observations (colours), with $\lambda_{\rm Edd}$ computed with the $L_{2\,\rm keV}$+K13 prescription and each object's SE mass. Coloured lines show the per-object relations; the black line is a single fit to all the points shown, analogous to the population analysis of Sect.~\ref{sec:wholesample}; the dashed line has the population-mean slope $\langle a_\lambda\rangle$. A relation that per-object is steep, becomes washed out and the slope even goes negative when the points are fit all together.}
\label{fig: envelope}
\end{figure*}

The two analyses start from the same kind of objects but lead to different results. The population regression recovers slopes that are flat and depend on the mass indicator, while the multi-epoch analysis recovers a positive, significant slope. Figure~\ref{fig: envelope} illustrates how the two are connected, using the five objects with the largest number of epochs. Each object traces its own positive relation, but the five tracks are offset from one another, and a single regression through all the points gives a slope of $-0.11\pm0.02$. If $\Gamma$ were set by $\lambda_{\rm Edd}$ alone and masses were known perfectly, the tracks would lie on a single line. They do not, and the main reason behind it we argue is the mass noise, which shifts each track along the $x$-axis by an amount comparable to the whole range of $\lambda_{\rm Edd}$ in the sample.  This is reminiscent of Simpson's paradox \citep{Simpson51}, in which a relation that holds within each group vanishes or reverses when the groups are pooled. Here the relation within objects is lost in the population because of errors in the independent variable.


\section{Discussion}
\label{sec:discussion}

\subsection{The Eddington ratio does not set the photon index}
\label{sec:disc_gamma}

Our multi-epoch analysis shows that the $\Gamma$--$\lambda_{\rm Edd}$ coupling is real, but its amplitude, $\langle a_\lambda\rangle\approx0.20-0.25$, is modest compared with the observed diversity of photon indices. In our sample, $\langle\Gamma\rangle=2.02$ with an observed dispersion of 0.27, of which $0.148\pm0.003$ is intrinsic once the measurement uncertainties (median $\sigma_\Gamma=0.25$) are accounted for. If the intrinsic spread of $\log\lambda_{\rm Edd}$ among SDSS quasars is $\lesssim0.15$\,dex \citep{Risaliti26}, the Eddington ratio accounts for a dispersion of only a fifth of the intrinsic one. Even across the full range of $\lambda_{\rm Edd}$ commonly quoted for type-1 AGN, $10^{-2}$--1, the relation would produce a change of 0.4--0.5 in $\Gamma$. The diversity of $\Gamma$ among unobscured quasars must therefore be driven by other parameters, such as the geometry, heating, or optical depth of the corona.\\
A direct consequence concerns the predictive power of $\Gamma$ for an individual source. Even at fixed mass, the relation is too shallow, and too diverse from object to object, to be inverted. 
With $\langle a_\lambda\rangle\approx0.2$ and a scatter of $\Gamma$ at fixed $\lambda_{\rm Edd}$ of 0.09--0.19 (Table~\ref{tab:wholesample_summary}), a single measurement of $\Gamma$ with typical photometric uncertainties constrains $\log\lambda_{\rm Edd}$ only to within $\approx1$\,dex, even with hypothetically perfect black-hole masses. What our results do support is a differential statement: when the same object is monitored over time, a steeper $\Gamma$ comes with brighter states; when monochromatic luminosities are translated in $\lambda_{\rm Edd}$ values, this means a more efficiently accreting object.

\subsection{Implications for X-ray weak AGN at high redshift}
\label{sec:disc_lrd}
These considerations apply directly to the X-ray weak AGN discovered by JWST, including the little red dots. In some studies, their X-ray weakness is attributed to high-Eddington accretion producing very steep X-ray spectra. With the slopes we measure, $\langle a_\lambda\rangle\approx0.2$--0.25, steepening $\Gamma$ from the typical 1.9 to $\gtrsim$\,3 would require an increase of $\lambda_{\rm Edd}$ by four to five orders of magnitude. This estimate extrapolates our relation far beyond the range we sample (most of our objects have $\log\lambda_{\rm Edd}\approx-1.5$ to 0), and the behaviour of the corona in the super-Eddington regime may differ. However, the available data at $\lambda_{\rm Edd}\gtrsim1$ show no sign of systematic steepening: highly accreting quasars span $\Gamma=1.3$--2.5 and show no significant correlation between $\Gamma$ and $\lambda_{\rm Edd}$ \citep{Laurenti24}. \\
Luminous quasars at $z>6$ (e.g., the HYPERION sample) provide a complementary test. \cite{Zappacosta23} found that they have steep X-ray spectra, with an average $\Gamma\approx2.4$, significantly steeper than the $\Gamma\approx1.8$--2 of lower-redshift quasars with comparable Eddington ratios, such as the hyper-luminous WISSH quasars at $z=2$--3 \citep{Zappacosta20}.\\
The difference of $\approx0.5$ in $\Gamma$ would require, with our slope, a difference of more than two orders of magnitude in $\lambda_{\rm Edd}$, and must therefore be driven by other parameters. \cite{Tortosa24} found that the steepness of the X-ray spectra of these quasars correlates with the strength of their UV disc winds, and noted that it may partly reflect a low-energy cutoff of the coronal emission rather than a steep power law. Although a steeper $\Gamma$--$\lambda_{\rm Edd}$ relation for super-Eddington sources has also been invoked to explain such spectra \citep[e.g.,][]{Huang20}, our results suggest that high Eddington ratios alone are unlikely to explain very steep X-ray spectra, and that alternative explanations, such as extreme obscuration, disc winds, or an intrinsically weak or undeveloped corona \citep{Maiolino25, Sacchi25, Madau24}, are favoured.

\subsection{Is the Eddington ratio the right tracer of the accretion state?}
\label{sec:disc_niccolai}
Our result can be read in two ways: the corona may genuinely respond only weakly to changes in the accretion state, or $\lambda_{\rm Edd}$, as computed from the observed continuum through standard bolometric corrections, may not faithfully trace those changes. The quantity we measure robustly is the response of $\Gamma$ to the X-ray continuum. Its translation into a slope in $\lambda_{\rm Edd}$ depends on how the bolometric output of an individual object changes when its continuum varies, i.e.\ on $\kappa$ (Sect.~\ref{sec:conversion}).
There are reasons to suspect that the continuum variations of individual quasars do not simply reflect proportional changes in the bolometric output. Quasars vary significantly on timescales of weeks to years, much shorter than the viscous timescale of a standard disc \citep{Lawrence18}, and this variability is often attributed to reprocessing of the variable X-ray emission or to local fluctuations of the disc rather than to changes of $\dot M$ \citep[e.g.,][]{Sun20}. Moreover, in individual AGN the broad emission lines respond less than proportionally to continuum variations \citep[the intrinsic Baldwin effect; e.g.,][]{Pogge92, Rakic17}, and the assumptions of the standard thin disc are expected to break down at high accretion rates, where radiation-pressure-driven winds may limit the temperature of the inner disc \citep{Laor14, Abramowicz13}. Preliminary results on the response of the broad lines to continuum variations in large samples of SDSS quasars suggest that, at fixed mass, the bolometric output may change much less than the observed continuum (Niccolai et al., in prep.). In this case, the variations of $L_{\rm X}$ and $\Gamma$ that we observe would trace changes in the structure of the inner accretion flow rather than in $\lambda_{\rm Edd}$.

\subsection{Comparison with state-dependent coupling}
\label{sec:disc_states}

Several recent studies found that the disc--corona coupling depends on the accretion state \citep[for a review]{RicciTrakhtenbrot23}. Changing-look AGN undergo optical-state transitions at a median $\log\lambda_{\rm Edd}\approx-2.0$ \citep{Jana25}, matching the hard-to-soft transition in X-ray binaries \citep[e.g.,][]{NodaDone18}, and show a V-shaped $\Gamma$--$\lambda_{\rm Edd}$ relation with a break at $\log\lambda_{\rm Edd}=-2.47\pm0.09$, below which the soft excess also disappears \citep{Jana26}. \cite{Middei26} reached a parallel conclusion from the disc side: in six years of simultaneous UV and X-ray monitoring of the Seyfert ESO\,511-G030, the UV and X-ray emission follow the non-linear $L_{\rm X}$--$L_{\rm UV}$ relation of quasars above $\lambda_{\rm Edd}\approx0.01$, and decouple below it, as expected if the inner disc evaporates into a hot, radiatively inefficient flow. Our sample lies almost entirely above this threshold, and our positive slope is consistent with the disc-dominated branch of the relation.

\section{Conclusions}
\label{sec:conclusions}

We have presented a new analysis of the $\Gamma$--$\lambda_{\rm Edd}$ relation using a homogeneous sample of 7835 type-1 quasars from SDSS DR16 and 4XMM--DR14. These are our main findings:

\begin{enumerate}
\item Fitting $\Gamma$ against $\log\lambda_{\rm Edd}$ with one point per object and single-epoch masses, the recovered slope depends on both the bolometric correction and the mass indicator, from $a_\lambda\approx0.10$--0.29 for H$\beta$ masses to $-0.11$--$-0.03$ for C\,{\sc iv} masses. The slopes are mostly flatter than literature values, and their disagreement is not reduced by a sample an order of magnitude larger than previous ones.

\item These results are explained by the composition of the $\log\lambda_{\rm Edd}$ axis. The systematic uncertainty of single-epoch masses ($\sim$\,0.4\,dex) is much larger than the intrinsic spread of $\lambda_{\rm Edd}$ in SDSS quasars, so that most of the variance of the $x$-axis is mass noise. Forward-modelled mock samples that reproduce our selection show that, in these conditions, any intrinsic slope up to 0.4 is recovered as $|a_\lambda|\lesssim0.03$(Appendix~\ref{app:mock}). The non-zero slopes we measure therefore reflect effects specific to each subsample, such as mass errors correlated with $\Gamma$ through the line width or the contribution of the soft excess to the photometric $\Gamma$ at low redshift, rather than the $\Gamma$--$\lambda_{\rm Edd}$ relation itself. 

\item Using 893 objects with multiple \xmm{} observations, we removed the black-hole mass from the $x$-axis and measured the response of $\Gamma$ to changes in the X-ray luminosity of each object. Stacking the per-object slopes with a hierarchical model gives $\langle a_X\rangle=0.35\pm0.03$, significantly different from zero, and stable against the minimum number of epochs and the variability selection. Individual objects rarely constrain their own slope, since their epoch-to-epoch variations of $\Gamma$ are consistent with the photometric uncertainties: the signal is a property of the population. When only precisely measured photon indices are used ($\Gamma/\sigma_\Gamma\gtrsim10$), the slope steepens and reaches a plateau at $\langle a_X\rangle\approx0.43$.

\item Converted through the $L_{\rm X}$--$L_{\rm UV}$ relation and a constant bolometric correction, the response corresponds to $\langle a_\lambda\rangle=0.21\pm0.02$, or $\approx0.25$ for precise photon indices. The conversion assumes that population relations describe the response of individual objects; plausible per-object values of the $L_{\rm X}$--$L_{\rm UV}$ slope and of the bolometric scaling, including a standard thin disc, give 0.14--0.26. 

\item The coupling is real but shallow. Across quasars, the narrow range of $\lambda_{\rm Edd}$ accounts for only a fifth of the dispersion in $\Gamma$ values: the Eddington ratio is not the main driver of the diversity of photon indices in unobscured AGN.

\item As a consequence, $\Gamma$ cannot be used to infer $\lambda_{\rm Edd}$ for individual sources, even with a perfectly known mass. Nor can high Eddington ratios alone explain very steep X-ray spectra in X-ray weak AGN at high redshift, such as the little red dots: with the slopes we measure, steepening $\Gamma$ from 1.9 to $\gtrsim3$ would require an increase of $\lambda_{\rm Edd}$ by four to five orders of magnitude.
\end{enumerate}

Two natural next steps to the present work are foreseen. First, a spectroscopic determination of $\Gamma$ for the whole multi-epoch sample would reduce the uncertainties that currently limit the per-object constraints, as suggested by the steepening of the slope for precise photometric measurements. Second, dedicated monitoring campaigns of individual sources, with high cadence and simultaneous X-ray and optical/UV coverage, could measure directly how the corona, the disc, and the bolometric output of a single object respond to changes in the accretion rate, which the serendipitous \xmm{} archive cannot probe.

\FloatBarrier   
\begin{acknowledgements}
MS acknowledges support through the European Space Agency (ESA) Research Fellowship Programme in Space Science. BT acknowledge support by European Union’s HE ERC Starting Grant No. 101040227 - WINGS. Views and opinions expressed are however those of the authors only and do not necessarily reflect those of the European Union or the European Research Council Executive Agency. Neither the European Union nor the granting authority can be held responsible for them. AS is supported by the national doctoral scholarship from the Agencia Nacional
de Investigación y Desarrollo (ANID), folio de postulación 21221788. AS personally acknowledges Olesya and Kirill Kuchay for their unwavering support. MR has been supported by the Polish National Agency for Academic Exchange (Bekker grant BPN/BEK/2024/1/00298).

\end{acknowledgements}

\bibliographystyle{aa}
\bibliography{bibl_cleaned}

\appendix
\section{Bolometric corrections: details}
\label{app:kbol}

This appendix describes the bolometric corrections summarised in Table~\ref{tab:kbol}. All optical/UV corrections use the continuum luminosities $\nu L_\nu(\lambda)$ tabulated in DR16Q \citep{Wu22}; when the anchor wavelength is not covered by the SDSS spectrum of an object, the prescription is not applied to it (Table~\ref{tab:wholesample_summary}).
\paragraph{K13.} \cite{Krawczyk13} constructed mean SED templates from 119\,652 SDSS broad-line quasars at $0.064<z<5.46$, using mid-IR data from Spitzer and WISE, near-IR from 2MASS and UKIDSS, optical from SDSS, and UV from GALEX. Integrating the mean SED from 1\,$\mu$m to 2\,keV, to avoid double counting of IR and hard X-rays, they obtained
\begin{equation}
  L_{\rm bol}^{\rm K13} = (2.75\pm 0.40)\times \nu L_\nu(2500\,\AA).
  \label{eq:bc_krawczyk}
\end{equation}
We use $L_{2500}$ from DR16Q when available (7456 objects) and otherwise derive it from $L_{3000}$ with the UV slope given below (255 objects).
\paragraph{K13$_{5100}$.} \cite{Krawczyk13} also provide a constant correction at 5100\,\AA,
\begin{equation}
  L_{\rm bol}^{\rm K13,5100} = (4.33\pm1.29)\times \nu L_\nu(5100\,\AA),
  \label{eq:bc_saccheo}
\end{equation}
integrated over the same range; including the 2--10\,keV band raises it to $4.80\pm1.54$. \cite{Saccheo23} found the same value, $K_{5100}=4.8\pm0.8$, for the hyper-luminous WISSH quasars ($L_{\rm bol}\gtrsim10^{47}$\,erg\,s$^{-1}$), showing that the correction does not depend on luminosity up to the brightest sources. $L_{5100}$ is available only at $z\lesssim0.8$.
\paragraph{D20.} From a sample of $\sim$\,1000 type-1 AGN with full SED coverage, \cite{Duras20} derived
\begin{equation}
  L_{\rm bol}^{\rm D20} = (5.18\pm0.12)\times\nu L_\nu(4400\,\AA),
  \label{eq:bc_duras}
\end{equation}
independent of $L_{\rm bol}$, $M_{\rm BH}$ and $\lambda_{\rm Edd}$, with an intrinsic scatter of $\sim$\,0.27\,dex. We obtain $L_{4400}$ by interpolating between $L_{3000}$ and $L_{5100}$ when both are available, and otherwise by extrapolating from the available one.
\paragraph{N19.} \citet{Netzer19} combined thin accretion-disc models, covering a range of black-hole masses, spins and accretion rates, with empirical X-ray properties, and approximated the resulting corrections with power laws, $K=c\,[\nu L_\nu(\lambda)/10^{42}\,{\rm erg\,s^{-1}}]^{d}$. At 3000\,\AA,
\begin{equation}
  L_{\rm bol}^{\rm N19} = 25\left(\frac{\nu L_\nu(3000\,\AA)}{10^{42}\,\rm erg\,s^{-1}}\right)^{-0.2}\nu L_\nu(3000\,\AA),
  \label{eq:lbol_netzer_l3000}
\end{equation}
i.e.\ $\kappa=0.8$ (at 1400\,\AA, $c=7$ and $d=-0.1$). The relations are eye-fitted to models with spin $a=0.7$ ($\eta\approx0.1$) and a mean inclination of $\sim$\,56$^\circ$. N19 does not quote a scatter: the full range of the model corrections is comparable to the range of radiative efficiencies (a factor of $\sim$\,8), and reduces to $\pm$0.05--0.35\,dex when $M_{\rm BH}$ is known. N19 stresses that such corrections approximate the mean properties of the population but can be wrong by an order of magnitude or more for individual sources. In particular, the negative exponent reflects the differences between objects of different mass and spin: at fixed $M_{\rm BH}$ and spin, the model corrections increase with the accretion rate (their Fig.~1), i.e.\ $\kappa>1$. We therefore use N19 only in the population analysis, and not to convert the per-object slopes of Sect.~\ref{sec:conversion}.
\paragraph{SS.} In the $\nu^{1/3}$ part of the spectrum of a standard thin disc \citep{Shakura73}, the monochromatic luminosity is $4\pi D_L^2F_\nu=f(i)\,[M_8\dot M]^{2/3}(\lambda/5100\,\AA)^{-1/3}$, with $M_8=M_{\rm BH}/10^8\,M_\odot$, $\dot M$ in $M_\odot$\,yr$^{-1}$, and $f(i)=f_0\cos i\,(1+b\cos i)/(1+b)$, where $b\simeq2$ describes limb darkening and $f_0\simeq1.2\times10^{30}$\,erg\,s$^{-1}$\,Hz$^{-1}$ \citep{Collin02, Davis11, Netzer14}. With $L_{\rm bol}=\eta\dot Mc^2$ this gives
\begin{equation}
  \log L_{\rm bol}^{\rm SS} = \frac{3}{2}\log\nu L_\nu(\lambda) - \log M_{\rm BH} + C_{\rm SS}(\lambda),
  \label{eq:lbol_ss}
\end{equation}
where we adopt $\eta=0.1$ and $\cos i=0.8$, so that $C_{\rm SS}=-13.28$, $-13.74$, $-13.90$ and $-14.24$ at 5100, 3000, 2500 and 1700\,\AA. These choices shift $\log\lambda_{\rm Edd}^{\rm SS}$ by a constant and do not affect the slopes. The approximation is most accurate at long wavelengths, with deviations increasing with $M_{\rm BH}$ and towards shorter wavelengths \citep{Netzer14}; for each object we therefore use the longest available anchor among 5100, 3000, 2500 and 1700\,\AA\ (2030, 5208, 501 and 119 objects, respectively). Unlike the empirical corrections, $L_{\rm bol}^{\rm SS}$ depends on the black-hole mass, so that $\log\lambda_{\rm Edd}^{\rm SS}=1.5\log\nu L_\nu-2\log M_{\rm BH}+{\rm const}$ and the mass uncertainty enters $\lambda_{\rm Edd}$ twice. At fixed mass, $\kappa=3/2$.
\paragraph{X-ray route.} For the prescriptions labelled $L_{2\,\rm keV}$+, $L_{2500}$ is derived from $L_{\rm X}$ by inverting the $L_{\rm X}$--$L_{\rm UV}$ relation,
\begin{equation}
  \log L_{2500} = \frac{\log L_{\rm X} - \beta}{\gamma},
  \label{eq:alphaox_inv}
\end{equation}
with $\gamma=0.591$ and $\beta=8.627$ \citep{Lusso20, Benetti25}. The slope is measured by fitting the flux--flux relation in narrow redshift bins, which makes it independent of cosmology; the normalisation is fixed by cross-calibrating the quasar Hubble diagram with type Ia supernovae. Only $\gamma$ enters the slopes derived in this work, while $\beta$ affects only the absolute values. The observed dispersion of the relation is $\approx0.15$\,dex \citep{Risaliti26b}, most of which is due to X-ray variability and inclination \citep{Signorini24}. For corrections anchored at other wavelengths, $L_{2500}$ is converted assuming a UV slope $\alpha_\nu=-0.44$ \citep[$L_\nu\propto\nu^{\alpha_\nu}$;][]{VandenBerk01}.; a per-object spread $\sigma_{\alpha_\nu}\simeq0.5$ adds $\sigma_{\alpha_\nu}|\log(\lambda/2500\,\AA)|\approx0.04$, 0.12 and 0.15\,dex at 3000, 4400 and 5100\,\AA\ to the scatter of $L_{\rm bol}$.

\section{Cut on $\Gamma$}
\label{app:gamma}
\begin{figure}[!htbp]
\centering
\includegraphics[width=0.8\linewidth,clip]{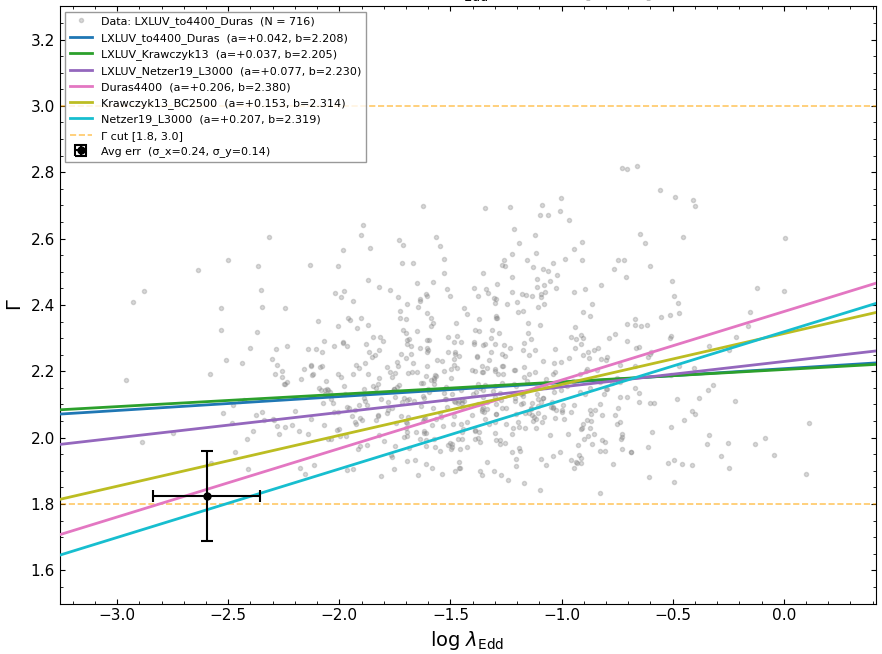}
\includegraphics[width=0.8\linewidth,clip]{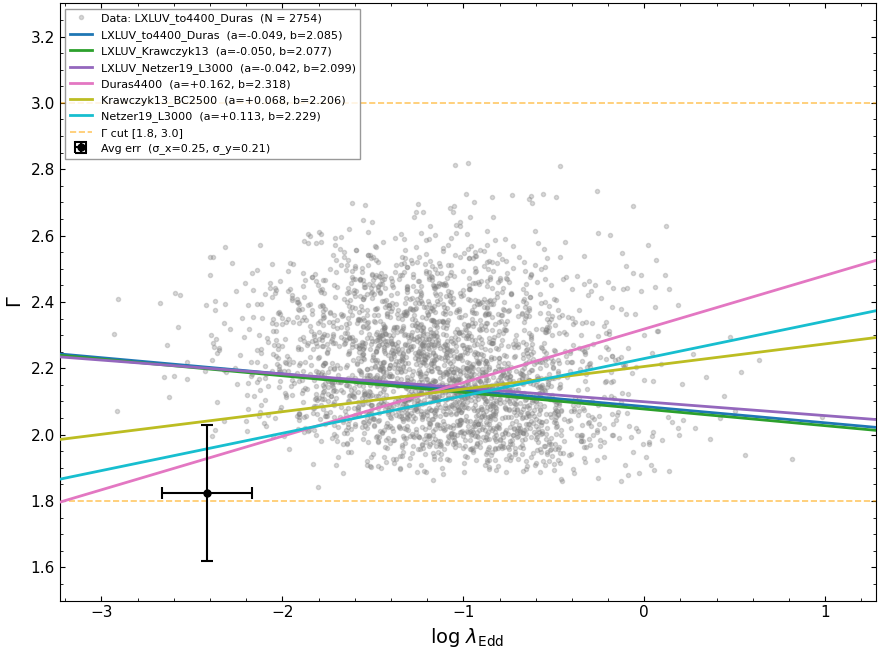}
\includegraphics[width=0.8\linewidth,clip]{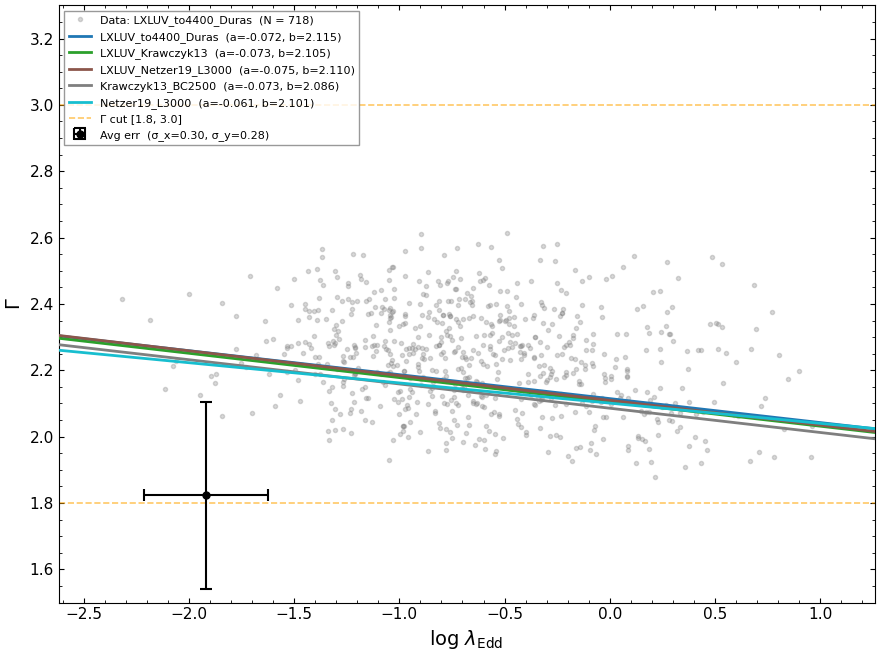}
\caption{As Fig.~\ref{fig:wholesample_Hb_MgII_CIV}, for $\Gamma-\Delta\Gamma>1.8$. }
\label{fig:wholesample_Hb_MgII_CIV_g18}
\end{figure}

We repeat the analyses requiring $\Gamma-\Delta\Gamma>1.8$ instead of 1.5. The single-epoch results (Fig.~\ref{fig:wholesample_Hb_MgII_CIV_g18}) are fully consistent with those of Sect.~\ref{sec:wholesample}, with a smaller statistic. 

\section{Monochromatic vs.\ integrated X-ray luminosity: simulations}
\label{app:mock_l210}
The integrated 2--10\,keV luminosity is not measured independently of the photon index: 
\begin{equation}
  L_{2-10} = \nu L_\nu(2\,\rm keV)\;2^{\Gamma-2}\,
    \frac{10^{2-\Gamma}-2^{2-\Gamma}}{2-\Gamma},
  \label{eq:l210_integral}
\end{equation}
At fixed $L_{2\,\rm keV}$, a steeper spectrum gives a lower $L_{2-10}$, with ${\rm d}\log L_{2-10}/{\rm d}\Gamma\approx-0.3$ over the range of $\Gamma$ of our sample. A measurement error $\delta\Gamma$ therefore also shifts $\log L_{2-10}$ by $\approx-0.3\,\delta\Gamma$, and the errors on the two axes of a $\Gamma$--$\log L_{2-10}$ relation are correlated. The monochromatic luminosity at 2\,keV does not have this problem, since its uncertainty is independent of that on $\Gamma$. To quantify the effect on the slope of a $\Gamma$--luminosity relation, we performed a set of simulations that reproduce our population analysis.

\begin{figure}[!htbp]
\centering
\includegraphics[width=0.7\linewidth,clip]{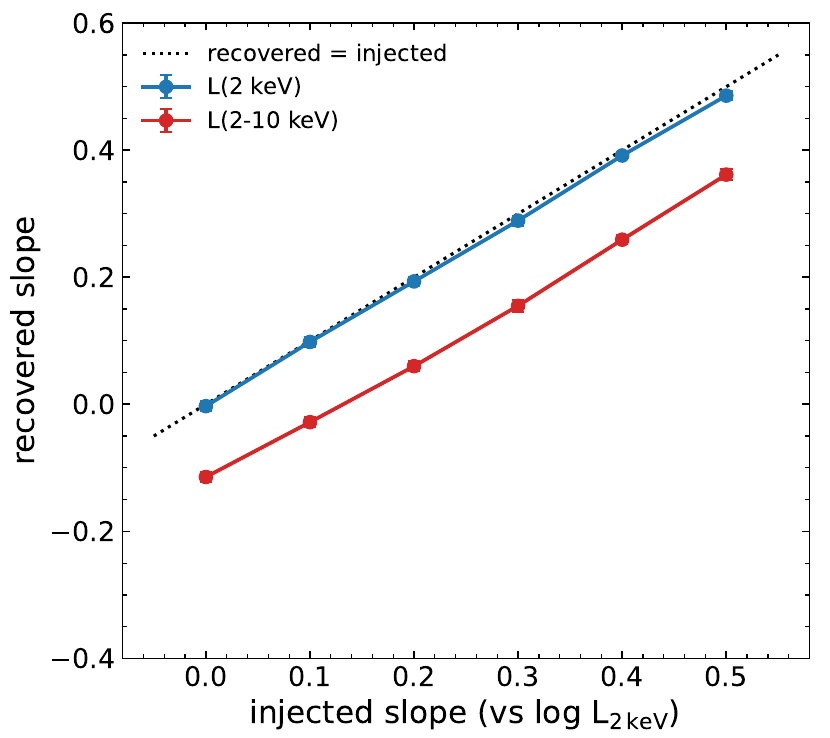}
\caption{Slope of the $\Gamma$--luminosity relation recovered from simulated samples as a function of the injected slope of the $\Gamma$--$\log L_{2\,\rm keV}$ relation. Blue: fit against the monochromatic luminosity at 2\,keV; red: fit against the integrated 2--10\,keV luminosity, computed from the observed $L_{2\,\rm keV}$ and $\Gamma$. Points and error bars are the mean and standard deviation over the realisations; the dotted line marks equality between recovered and injected slope.}
\label{fig:mock_l210}
\end{figure}

For each simulated sample we drew 6000 values of $\log L_{2\,\rm keV}$ from the observed distribution in our sample, and assigned a photon index $\Gamma = 2.0 + a_{\rm in}\,(\log L_{2\,\rm keV} - \log L_{\rm piv}) + \epsilon$, where $\log L_{\rm piv}$ is the median of the distribution, $a_{\rm in}$ is the injected slope, and $\epsilon$ is an intrinsic scatter of 0.15, drawn from a Gaussian distribution. We then added measurement noise to $\log L_{2\,\rm keV}$ and $\Gamma$, drawing the uncertainties from the observed distributions of $\sigma(\log L_{2\,\rm keV})$ and $\sigma_\Gamma$, independently of each other. From the noisy $L_{2\,\rm keV}$ and $\Gamma$ we computed $L_{2-10}$ with the above equation, as for the real data, and propagated the uncertainty on $\log L_{2-10}$ from those on $\log L_{2\,\rm keV}$ and $\Gamma$. Finally, we fitted $\Gamma$ against $\log L_{2\,\rm keV}$ and against $\log L_{2-10}$ with the same effective-variance likelihood used throughout the paper, which includes the uncertainties on both axes but not their covariance. We repeated the procedure for injected slopes $a_{\rm in}=0$--0.5, with 20 realisations for each value.
The results are shown in Fig.~\ref{fig:mock_l210}. When $\Gamma$ is fitted against the monochromatic luminosity, the injected slope is recovered over the whole range. When $\Gamma$ is fitted against $\log L_{2-10}$, instead, the recovered slope is systematically lower by $\approx0.11$--0.14. The bias arises because the correlated errors scatter each point along a direction of negative slope in the $\Gamma$--$\log L_{2-10}$ plane, and because the intrinsic scatter of $\Gamma$ at fixed luminosity also propagates into $\log L_{2-10}$. This shows that the integrated 2--10\,keV luminosity cannot be used as the independent variable in a $\Gamma$--luminosity relation without an explicit treatment of the covariance, and supports our choice of the monochromatic luminosity at 2\,keV.

\section{Photometric and spectroscopic photon indices}
\label{app:spec}

\begin{figure}[!htbp]
\centering
\includegraphics[width=0.7\linewidth,clip]{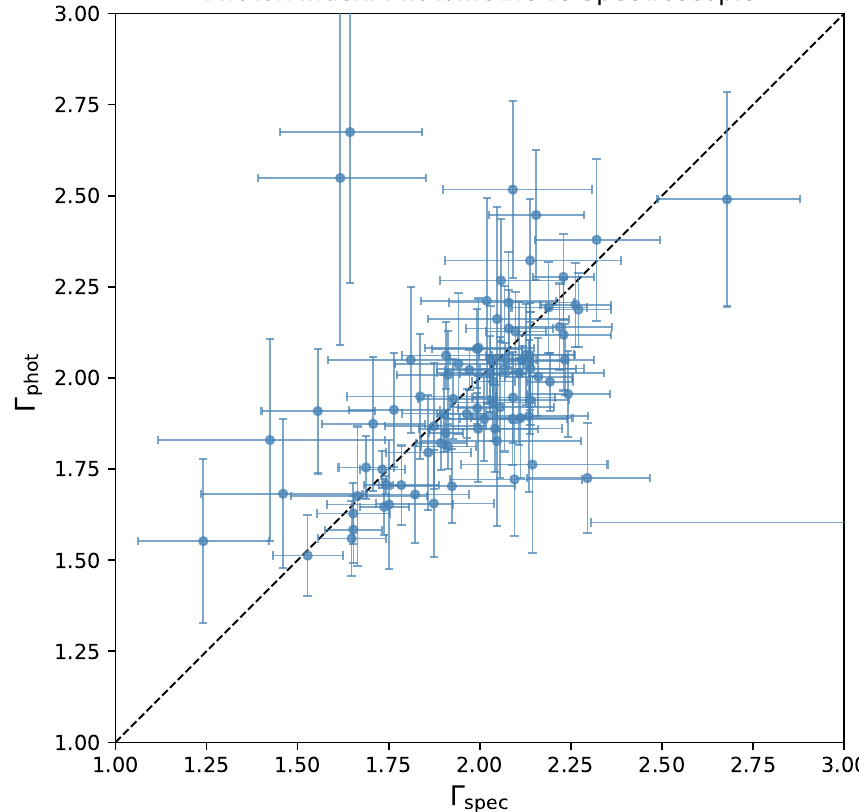}
\caption{Photometric vs.\ spectroscopic photon index for 90 randomly selected objects of our sample. The dashed line is the one-to-one relation. The comparison shows no systematic offset. No object required obscuration for their best fit. Spectroscopy reduces uncertainties by up to 40\%.}
\label{fig:gamma_phot_spec}
\end{figure}

The photon indices used throughout this work are derived from the \xmm{} photometry, i.e.\ from the ratio of the fluxes in the 1--2 and 2--4.5\,keV bands. To test their reliability, we extracted and fitted the \xmm{} spectra of 90 randomly selected objects of our sample, and compared the resulting photon index $\Gamma_{\rm spec}$ with the photometric one, $\Gamma_{\rm phot}$ (Fig.~\ref{fig:gamma_phot_spec}). The two estimates are consistent: a linear fit is consistent with the one-to-one relation, and the mean offset, $\langle\Gamma_{\rm phot}-\Gamma_{\rm spec}\rangle=-0.04$, is small compared with the typical uncertainties. None of the spectra requires intrinsic absorption. We are therefore confident that the photometric photon indices provide an unbiased estimate of the spectral slope, and that they can be used for the analyses presented in this paper. The comparison also shows a non-negligible scatter around the one-to-one relation, including a few objects with large discrepancies. The mean uncertainty on the photon index decreases from 0.21 for the photometric to 0.13 for the spectroscopic estimates. Since the per-object analysis of Sect.~\ref{sec:multiepoch} is limited by the precision of $\Gamma$ (Sects.~\ref{sec:perobject} and \ref{sec:snr}), a spectroscopic analysis of the whole sample, which we plan for a future work, will allow a more precise measurement of the response of $\Gamma$ to changes in luminosity.

\section{Fitting framework}
\label{app:fitting}

We model the relation as $\Gamma = a_{\lambda}\,\log\lambda_{\rm Edd} + b_\lambda$, with an additional intrinsic scatter $\sigma_{\rm int}$ in the $\Gamma$ direction, and use the effective-variance likelihood
\begin{equation}
  \mathcal{L} \propto
    \prod_{i}
    \frac{1}{s_i}
    \exp\!\left[{-\frac{(\Gamma_i - a_\lambda\,x_i - b_\lambda)^2}{2\,s_i^2}}\right],
  \label{eq:effvar_lkh}
\end{equation}
where $x_i$\,=\,$\log\lambda_{{\rm Edd},i}$ and $s_i^2$\,=\,$\sigma_{\Gamma,i}^2$\,+\,$a_\lambda^2\,\sigma_{x,i}^2$\,+\,$\sigma_{\rm int}^2$. Here $\sigma_{\Gamma,i}$ is the uncertainty on the photon index and $\sigma_{x,i}^2=\sigma_{M,i}^2+\sigma_{L,i}^2$ the uncertainty on $\log\lambda_{\rm Edd}$, with $\sigma_{M,i}$ the formal uncertainty on $\log M_{\rm BH}$ and $\sigma_{L,i}$ that on $\log L_{\rm bol}$ propagated from the anchor luminosity.

We sample the slope through the angle $\phi=\arctan(a_\lambda)$ with a uniform prior on $\phi\in(-\pi/2,\pi/2)$, which gives equal prior weight to every direction in the plane, whereas a uniform prior on $a_\lambda$ concentrates most of the prior volume on nearly vertical lines \citep{Hogg10}. 
The posterior is sampled with \texttt{emcee} \citep{emceepaper}, using 32 walkers, 1500 steps, and 50\% burn-in; the integrated autocorrelation time is $\tau\approx29$ steps (median over all fits; maximum 36), so that the chains after burn-in are at least 20 times $\tau$ and each posterior contains $\gtrsim800$ independent samples. We subtract the sample means of $x$ and $\Gamma$ before fitting, to reduce the correlation between slope and intercept, and restore them when reporting the intercept.

In Section \ref{sec:multiepoch} we use the same method, but excluding the intrinsic scatter term. We sample $\phi_j=\arctan(a_{X,j})$ with a uniform prior on $(-\pi/2,\pi/2)$ using \texttt{emcee} with 24 walkers and 2000 steps, discarding the first 25\%. The integrated autocorrelation time is $\tau\approx30$ steps (median; 90th percentile 35), so that the chains are about $50\tau$ long and each posterior contains $\approx1200$ independent samples. Here the choice of the prior matters: with two or three epochs the likelihood constrains the slope only weakly, and the posterior shape is partly set by the prior. The uniform prior in $\phi$ does not favour any direction of the line, and it is also the prior required by the stacking step, which infers the population distribution of $\phi$ from per-object samples drawn under a flat prior in the same variable (Sect.~\ref{sec:stacking}).

The per-object posteriors are combined with \texttt{PosteriorStacker} \citep{Baronchelli20} in the same variable $\phi=\arctan(a_X)$ in which they were sampled, which is consistent with the flat prior used for the individual fits. The parent distribution is modelled as a Gaussian in $\phi$ with mean $\mu_\phi$ and dispersion $\sigma_\phi$, and sampled with the nested-sampling code \texttt{UltraNest} \citep{Buchner21}. The population-mean slope is $\langle a_X\rangle=\tan\mu_\phi$, and its uncertainty is half the 16th--84th percentile range of $\tan\mu_\phi$ over the posterior. The dispersion of the per-object slopes quoted in Table~\ref{tab:multiepoch_lx_alx} is obtained as $\sigma_a=(1+\langle a_X\rangle^2)\,\sigma_\phi$, from ${\rm d}a/{\rm d}\phi=1+a^2$; for the full sample, $\mu_\phi=0.34$ and $\sigma_\phi=0.25$.

Over the full sample, 120 of the 893 objects (13\%) have $a_{X,j}>0$ at 1$\sigma$ and 35 (4\%) at 2$\sigma$, while only 27 (3\%) and 2 have $a_{X,j}<0$ at the same significance. The asymmetry grows with the number of epochs: for the 45 objects with at least ten observations, 42\% and 13\% have positive slopes at 1 and 2$\sigma$, against 2\% and 2\% negative. The weakness of the individual constraints reflects the precision of the photometric photon indices (median $\sigma_\Gamma=0.20$ per epoch), which is comparable to or larger than the epoch-to-epoch scatter of $\Gamma$ ($\langle{\rm std}(\Gamma)\rangle=0.19$, median 0.16). Individual objects therefore rarely constrain their own slope, and the signal is a population property that emerges from stacking.

\begin{table*}[!tp]
\centering
\caption{Slopes recovered from the simulations of Appendix~\ref{app:mock} (mean and standard deviation over 20 realisations), for an intrinsic $\lambda_{\rm Edd}$ spread of 0.15\,dex and a virial-mass noise of 0.4\,dex.}
\label{tab:mock_selection}
\setlength{\tabcolsep}{5pt}
\begin{tabular}{ccccccc}
\hline\hline
 & \multicolumn{2}{c}{H$\beta$} & \multicolumn{2}{c}{Mg\,{\sc ii}} & \multicolumn{2}{c}{C\,{\sc iv}} \\
$a_{\rm in}$ & $L_{2\,\rm keV}$+K13 & K13 & $L_{2\,\rm keV}$+K13 & K13 & $L_{2\,\rm keV}$+K13 & K13 \\
\hline
0.0 & $-0.02\pm0.02$ & $-0.01\pm0.02$ & $-0.02\pm0.01$ & $-0.00\pm0.01$ & $-0.02\pm0.01$ & $-0.00\pm0.01$ \\
0.1 & $-0.01\pm0.02$ & $+0.01\pm0.02$ & $-0.01\pm0.00$ & $+0.01\pm0.01$ & $-0.02\pm0.01$ & $-0.00\pm0.01$ \\
0.2 & $-0.01\pm0.02$ & $+0.01\pm0.02$ & $-0.01\pm0.01$ & $+0.02\pm0.01$ & $-0.01\pm0.01$ & $+0.01\pm0.01$ \\
0.3 & $-0.00\pm0.01$ & $+0.03\pm0.02$ & $-0.00\pm0.01$ & $+0.02\pm0.01$ & $-0.00\pm0.01$ & $+0.02\pm0.01$ \\
0.4 & $-0.01\pm0.01$ & $+0.02\pm0.02$ & $+0.00\pm0.01$ & $+0.03\pm0.01$ & $-0.00\pm0.01$ & $+0.03\pm0.01$ \\
\hline
\end{tabular}
\end{table*}

\section{Simulations of the population analysis}
\label{app:mock}
We forward-modelled the population analysis of Sect.~\ref{sec:wholesample} to test whether an intrinsic $\Gamma$--$\lambda_{\rm Edd}$ relation survives the selection and the mass noise of our sample, and whether selection alone can produce the differences between the H$\beta$, Mg\,{\sc ii} and C\,{\sc iv} subsamples. All the observational ingredients were measured on the data:
\begin{itemize}
\item the redshift distribution of each subsample;
\item the effective X-ray and optical flux limits, defined as the 2nd percentile of $\log L_{\rm X}$ and $\log L_{2500}$ in 15 redshift bins;
\item the uncertainty on $\Gamma$ as a function of the distance of a source from the X-ray limit, $\log\sigma_\Gamma=-0.31-0.47\,(\log L_{\rm X}-\log L_{\rm X,lim})$, with a residual scatter of 0.24\,dex;
\item the uncertainties on $\log L_{\rm X}$ and the formal uncertainties on $\log M_{\rm BH}$, drawn from their observed distributions.
\end{itemize}
For each subsample we drew an intrinsic population with $\log\lambda_{\rm Edd}$ distributed as a Gaussian of width 0.15\,dex \citep{Risaliti26}, centred on the observed mean, and $\log M_{\rm BH}$ distributed as a Gaussian of width 0.3\,dex, centred 0.1\,dex below the observed mean. We computed $L_{2500}=K_{\rm K13}^{-1}\lambda_{\rm Edd}L_{\rm Edd}$ and $L_{\rm X}$ from the $L_{\rm X}$--$L_{\rm UV}$ relation with a scatter of 0.2\,dex, and assigned $\Gamma=2.0+a_{\rm in}(\log\lambda_{\rm Edd}-\langle\log\lambda_{\rm Edd}\rangle)$ with an intrinsic scatter of 0.15. After adding the measurement errors, we applied the flux limits and the selection $\Gamma - \Delta\Gamma>1.5$ and $\Gamma + \Delta\Gamma  < 3.5$, and kept as many objects as in the corresponding real subsample. Finally, we added a virial-mass noise of 0.4\,dex and fitted $\Gamma$ against $\log\lambda_{\rm Edd}$ computed with the K13 and $L_{2\,\rm keV}$+K13 routes, with the likelihood of Eq.~\ref{eq:effvar_lkh}.\\
The mock samples reproduce the observed properties of the three subsamples. The mean and dispersion of the observed masses are $\log M_{\rm BH}=8.09\pm0.49 M_{\odot}$, $8.65\pm0.49 M_{\odot}$ and $8.83\pm0.50 M_{\odot}$ in the H$\beta$, Mg\,{\sc ii} and C\,{\sc iv} mocks, against $8.17\pm0.47 M_{\odot}$, $8.73\pm0.44 M_{\odot}$ and $8.92\pm0.49 M_{\odot}$ in the data; those of the X-ray luminosity are $\log L_{\rm X}=25.91\pm0.26$, $26.37\pm0.26$ and $26.75\pm0.26$, against $25.96\pm0.34$, $26.39\pm0.35$ and $26.79\pm0.30$; and those of $\log\lambda_{\rm Edd}$ ($L_{2\,\rm keV}$+K13) are $-1.07\pm0.53$, $-0.85\pm0.54$ and $-0.38\pm0.54$, against $-1.08\pm0.53$, $-0.88\pm0.47$ and $-0.40\pm0.55$. In this configuration the flux limits remove only 3--6\% of the intrinsic population, since the observed luminosity distributions already lie close to the limits.\\
Table~\ref{tab:mock_selection} lists the slopes recovered from 20 realisations for injected slopes $a_{\rm in}=0$--0.4. In all three subsamples and for both routes, the recovered slopes are between $-0.02$ and $+0.03$, with realisation-to-realisation scatter of 0.01--0.02. For $a_{\rm in}=0$ the $L_{2\,\rm keV}$+K13 slopes are slightly negative ($-0.02$ in all subsamples): the $\Gamma$ selection removes low values of $\Gamma$ preferentially in faint sources, whose $\sigma_\Gamma$ is larger, and therefore induces a weak anti-correlation between $\Gamma$ and $L_{\rm X}$. Selection effects alone thus produce slopes consistent with zero in all subsamples, and cannot reproduce either the positive slopes measured in the H$\beta$ subsample (0.10--0.29) or the negative ones of the C\,{\sc iv} subsample ($-0.11$ to $-0.03$). As discussed in Sect.~\ref{sec:wholesample_MBHscatter}, these require effects that are not included in the simulations, such as mass errors correlated with $\Gamma$ or $\lambda_{\rm Edd}$, or a contribution of the soft excess to the measured $\Gamma$.

\section{Tests on the variability selection}
\label{app:varcut}

Table~\ref{tab:varcut} compares the 893 objects that pass the variability cut (Sect.~\ref{sec:multiepoch}) with the 808 that do not. The two groups differ slightly: the selected objects are at lower redshift (median $z=1.08$ against 1.59), have lower X-ray luminosity, more epochs, and slightly lower $M_{\rm BH}$ and $\lambda_{\rm Edd}$. Despite this, the stacked slope changes little with the threshold on ${\rm std}(\log L_{\rm X})/{\rm median}(\sigma_{\rm X})$: it is $0.33\pm0.03$ with no cut (1694 objects), and $0.35$--0.36 for thresholds between 1 and 2, with a stable population dispersion, $\sigma_\phi=0.25$--0.26. The slight increase is expected, since objects whose luminosity changes are dominated by noise dilute the slope.

Table~\ref{tab:bins_ml} reports the stacked slope in quartiles of black-hole mass and of mean X-ray luminosity. The slope varies between 0.26 and 0.45 across the mass quartiles, without a monotonic trend.  It decreases mildly with X-ray luminosity, from $0.47\pm0.08$ in the faintest to $0.30\pm0.05$ in the brightest quartile (a 2$\sigma$ difference). Since the selected objects are at lower redshift and luminosity than the rejected ones, this trend may contribute to the slight increase of the slope with the variability threshold. A possible contribution of the soft excess, which affects the photometric $\Gamma$ more strongly at low redshift, will be discussed in Sacchi et al.(in prep.).

\begin{table}[!tbp]
\centering
\caption{Top: comparison of the objects passing and failing the variability cut (medians and $p$-value of a two-sample KS test); $\lambda_{\rm Edd}$ from X+K13 and the fiducial DR16Q mass. Bottom: stacked slope ($N_{\rm min}=2$) as a function of the threshold on ${\rm std}(\log L_{\rm X})/{\rm median}(\sigma_{\rm X})$; the fiducial threshold is 1.}
\label{tab:varcut}
\begin{tabular}{lccc}
\hline\hline
Quantity & Pass & Fail & $p_{\rm KS}$ \\
\hline
$z$ & 1.08 & 1.59 & $2\times10^{-48}$ \\
$\langle\log L_{\rm X}\rangle$ & 26.36 & 26.55 & $6\times10^{-13}$ \\
$N_{\rm epochs}$ & 3 & 2 & $10^{-21}$ \\
$\log M_{\rm BH}$ & 8.67 & 8.80 & $2\times10^{-6}$ \\
$\log\lambda_{\rm Edd}$ & $-0.88$ & $-0.79$ & $8\times10^{-5}$ \\
\hline\hline
Threshold & $N$ & $\langle a_X\rangle$ & $\sigma_\phi$ \\
\hline
0.5 & 1248 & $0.33\pm0.03$ & 0.25 \\
1   &  893 & $0.35\pm0.03$ & 0.25 \\
1.5 &  616 & $0.35\pm0.03$ & 0.25 \\
2   &  448 & $0.36\pm0.03$ & 0.26 \\
\hline
\end{tabular}
\end{table}

\begin{table}[!tbp]
\centering
\caption{Stacked slopes ($N_{\rm min}=2$) in quartiles of the fiducial DR16Q black-hole mass and of the mean X-ray luminosity ($L_{\rm X}$ in erg\,s$^{-1}$\,Hz$^{-1}$).}
\label{tab:bins_ml}
\setlength{\tabcolsep}{4pt}
\begin{tabular}{lccc}
\hline\hline
Bin & $N$ & $\langle a_X\rangle$  \\
\hline
$7.14\le\log M_{\rm BH}<8.32$ & 222 & $0.45\pm0.07$  \\
$8.32\le\log M_{\rm BH}<8.67$ & 222 & $0.30\pm0.07$  \\
$8.67\le\log M_{\rm BH}<9.01$ & 222 & $0.45\pm0.04$  \\
$9.01\le\log M_{\rm BH}\le10.30$ & 223 & $0.28\pm0.05$  \\
\hline
$25.24\le\langle\log L_{\rm X}\rangle<26.08$ & 223 & $0.47\pm0.08$ \\
$26.08\le\langle\log L_{\rm X}\rangle<26.36$ & 223 & $0.35\pm0.07$  \\
$26.36\le\langle\log L_{\rm X}\rangle<26.65$ & 223 & $0.34\pm0.06$  \\
$26.65\le\langle\log L_{\rm X}\rangle\le28.11$ & 224 & $0.30\pm0.06$ \\
\hline
\end{tabular}
\end{table}

\end{document}